\documentclass[journal, twoside]{IEEEtran}
\usepackage{amsmath,amsfonts}
\usepackage{algorithmic}
\usepackage{algorithm}
\usepackage{array}
\usepackage[caption=false,font=normalsize,labelfont=sf,textfont=sf]{subfig}
\usepackage{textcomp}
\usepackage{stfloats}
\usepackage{url}
\usepackage{verbatim}
\usepackage{graphicx}
\usepackage{cite}
\usepackage{orcidlink}
\usepackage{tabularx}
\usepackage{booktabs}
\usepackage{multirow}
\usepackage{threeparttable}
\usepackage{ragged2e}
\usepackage{xcolor}
\usepackage{makecell}
\begin{document}

\title{NeuroSim V1.5: Improved Software Backbone for Benchmarking Compute-in-Memory Accelerators with Device and Circuit-level Non-idealities}

\author{James Read\orcidlink{0000-0003-0753-6257} ~\IEEEmembership{Graduate Student Member,~IEEE,} Ming-Yen Lee\orcidlink{0000-0002-5076-4645}~\IEEEmembership{Graduate Student Member, ~IEEE,} Wei-Hsing Huang\orcidlink{0009-0008-2405-2936}, Yuan-Chun Luo\orcidlink{0000-0001-5793-075X}~\IEEEmembership{Graduate Student Member,~IEEE,} Anni Lu\orcidlink{0000-0002-4415-0866}~\IEEEmembership{Graduate Student Member,~IEEE,} Shimeng Yu\orcidlink{0000-0002-0068-3652}~\IEEEmembership{Fellow,~IEEE,}

\thanks{This work is supported in part by PRISM, one of the SRC/DARPA JUMP 2.0 Centers. J. Read acknowledges support from the DoD's SCALE program.}%
\thanks{The authors are with the School of Electrical and Computer Engineering, Georgia Institute of Technology, Atlanta, GA, 30332 USA (emails: jread6@gatech.edu, shimeng.yu@ece.gatech.edu)}}

\markboth{IEEE TRANSACTIONS ON COMPUTER-AIDED DESIGN OF INTEGRATED CIRCUITS AND SYSTEMS}{READ \MakeLowercase{\textit{et al.}}: NEUROSIM V1.5: IMPROVED SOFTWARE BACKBONE FOR BENCHMARKING COMPUTE-IN-MEMORY ACCELERATORS}

\IEEEpubid{0000--0000/00\$00.00~\copyright~2026 IEEE}

\maketitle

\begin{abstract}
The exponential growth of artificial intelligence (AI) applications has strained conventional von Neumann architectures, where frequent data transfers between compute units and memory create significant energy and latency bottlenecks. Compute-in-Memory (CIM) addresses this challenge by performing multiply-accumulate (MAC) operations directly in memory arrays, substantially reducing data movement. As transformers increasingly dominate AI workloads, extending CIM frameworks to support these architectures becomes critical. In this work, we present NeuroSim V1.5, an integrated framework for co-optimizing accuracy and hardware efficiency in CIM accelerator design. Key contributions include: (1) a hybrid ACIM/DCIM architecture enabling transformer acceleration, with a case study comparing vision transformer (ViT) and CNN performance; (2) an integrated co-optimization framework combining TensorRT-based quantization, flexible noise modeling, and circuit-level PPA estimation; (3) expanded device support including non-volatile capacitive memories; and (4) up to 6.5$\times$ faster runtime through GPU-accelerated simulation. Our ViT case study reveals that ViTs exhibit lower noise tolerance and area efficiency than CNNs of comparable size, suggesting CNNs remain well-suited for vision applications on CIM hardware. The hybrid architecture provides the foundation for future transformer workloads including large language models. All versions are available open-source at https://github.com/neurosim/NeuroSim.
\end{abstract}


\section{Introduction}\label{introduction}
\IEEEPARstart{T}{he} rapid growth in AI model size has exposed the importance of energy-efficient processors. Memory access dominates power consumption in AI processing, consuming more than five times the energy of computation \cite{tripp_measuring_2024}. For large models, weights must be loaded periodically from DRAM, shifting the bottleneck from compute capability to memory bandwidth  \cite{williams_roofline_2009}.

\subsection{\textbf{Compute-in-Memory}}

Compute-in-memory (CIM) addresses this bottleneck by co-locating multiply-accumulate (MAC) operations with data storage. Digital CIM (DCIM) augments SRAM arrays with multipliers and adder trees \cite{chih_164_2021}. Analog CIM (ACIM) leverages memory cell physics for current-, charge-, or time-domain computation, where weights encode as conductance or capacitance, and column currents/charges are summed to yield MAC results \cite{yu_neuro-inspired_2018, verma_-memory_2019}. Various non-volatile memory (NVM) devices enable ACIM arrays, including RRAM \cite{jain_132_2019, 9755965, cheng-xin_xue_241_2019, qi_liu_332_2020}, PCM \cite{le_gallo_64-core_2023, vinay_joshi_accurate_2020}, FeFET \cite{mulaosmanovic_novel_2017, jerry_ferroelectric_2017, de_demonstration_2022}, Flash \cite{eFlash}, and nvCap \cite{10272016, luo_experimental_2021, 10237236}. SRAM-based implementations also exist for current-based \cite{sinangil_7-nm_2021, 8662392, 9062995} or charge-based \cite{valavi_64-tile_2019, lee_fully_2021} computation.

Both approaches present distinct trade-offs. ACIM with NVM can achieve over 100 TOPS/W \cite{chih_164_2021, qi_liu_332_2020} and high-density, enabling entire models to be stored on-chip. However, NVM devices suffer from limited write endurance, slow programming, and analog noise from device variations. DCIM offers robust digital computation, unlimited write endurance, and compatibility with advanced CMOS nodes but incurs higher area overhead and lower throughput. These complementary characteristics motivate heterogeneous architectures: ACIM excels for weight-stationary layers where weights are programmed once, while DCIM suits write-intensive operations like attention in transformers. As transformers increasingly dominate AI workloads---from vision transformers (ViTs) to large language models---extending CIM frameworks to support these architectures becomes critical.
\IEEEpubidadjcol

\subsection{\textbf{CIM Design Challenges}}

Optimizing CIM accelerators requires navigating a multi-dimensional design space spanning device characteristics (memory technology, precision, on/off ratio), circuit parameters (array dimensions, ADC precision), and system-level choices (quantization schemes, data mapping). The central challenge of ACIM is managing inherently noisy analog computations. Device-level fabrication imperfections cause device-to-device (D2D) variations and stuck-at-faults \cite{changhyuck_sung_effect_2018, wei_wu_methodology_2018}. Additionally, memory states drift over time \cite{ xiaoyu_sun_impact_2019}, and circuit-level thermal noise and process variations degrade ADC sensing margins \cite{luo_cross-layer_2024, 10237236}. At the system level, quantized partial sums compound errors through sequential layers \cite{matthew_spear_impact_2023, huang_hardware-aware_2023}. These multi-layered effects interact in complex ways, necessitating simulation frameworks that can rapidly evaluate both accuracy and hardware metrics across the design space. 

\subsection{\textbf{Existing Tool Limitations}}

Direct SPICE simulation is prohibitively slow for large networks, making broad parameter exploration impractical. Existing tools present different trade-offs: CrossSim \cite{xiao_tianyao_crosssim_2021} provides high-fidelity physical simulations but with high computational cost; AIHWKit \cite{rasch_flexible_2021} offers noise modeling primarily for training; CIMLoop \cite{andrulis_cimloop_2024} focuses on system architectures without accuracy modeling. No existing framework provides integrated accuracy-PPA co-optimization, particularly for trasnformers where self-attention mechanisms introduce unique challenges for CIM deployment.

In this paper, we present NeuroSim V1.5, an integrated simulation framework for systematic co-optimization of accuracy and hardware efficiency in CIM accelerator design. Unlike tools that address device physics or system performance in isolation, NeuroSim V1.5 combines behavioral simulation with detailed hardware analysis to enable comprehensive design space exploration. Our key contributions are:

\begin{enumerate}
    \item \textbf{Support for transformers with hybrid ACIM/DCIM architectures}: Our hybrid design augments NeuroSim's floorplan generator with DCIM tiles used for attention and LUTs for complex activation functions. We conduct a case study on ViT performance using this architecture and compare with CNNs.
    \item \textbf{An integrated co-optimization framework} combining TensorRT-based quantization, flexible noise modeling validated against SPICE/silicon data, and circuit-level PPA estimation—allowing rapid evaluation of accuracy-efficiency trade-offs across the full design space.
    \item \textbf{Expanded device support}, including emerging non-volatile capacitive memories (nvCap) and charge-domain compute methods.
    \item \textbf{Up to 6.5$\times$ runtime improvement} over NeuroSim V1.4 through optimized GPU-accelerated tensor operations.
\end{enumerate}

The remainder of this paper is organized as follows: Section \ref{background_info} provides technical background on CIM systems. Section \ref{methodology} details our framework architecture and noise modeling methodology. Section \ref{results} presents case studies including design space exploration, noise sensitivity analysis validated against SPICE and silicon measurements \cite{read_enabling_2023, yixin_multilevel_2024, luo_cross-layer_2024, 9755965}, PPA breakdown of Swin Transformer on ImageNet, and detailed error analysis comparing CNN and ViT architectures. Section \ref{conclusion} discusses future directions.

\section{Background Information}\label{background_info}
\subsection{\textbf{CIM Array Architecture and Operation}}

ACIM performs matrix computations by utilizing electrical properties of memory arrays. The fundamental building block is the crossbar array, where memory elements are placed at wordline (WL) and bitline (BL) intersections.

Crossbar arrays enable parallel MACs: input voltages applied to rows produce column currents/charges that naturally sum to implement dot products. This allows large-scale MAC operations in one analog step, vastly increasing throughput. Two primary analog computation methods have emerged.

Resistor-based arrays contain RRAM, PCM, Flash, FeFET or other resistive memory, which encode neural network weights as conductance states (G) and perform MAC operations in the current-domain:
\begin{equation}
I_{BL_i} = \sum_j G_{ij} V_{j}
\end{equation}
where $I_{BL_i}$ is the analog matrix-vector multiplication (MVM) output current on the $i^{th}$ bitline, and $j$ indexes across wordlines.

Capacitor-based arrays contain non-volatile capacitors (nvCap), which encode weights as programmable capacitance \cite{hur_nonvolatile_2022, kim_capacitive_2024}, performing MAC operations in the charge domain:
\begin{equation}
Q_{BL_i} = \sum_j C_{ij} V_{j}
\end{equation}

Alternative charge-based computation approaches use fixed capacitors with SRAM cells, where stored digital weight modulates charge accumulation via charge sharing \cite{9896828, jia_151_2021}.

DCIM \cite{chih_164_2021} uses digital multipliers and adder trees for in-memory integer MACs, offering robustness at the cost of higher area overhead compared to ACIM at the same technology node. The main advantage of DCIM is that it can be integrated in any CMOS technology, whereas non-volatile memories are mostly limited to legacy nodes.

\subsection{\textbf{Device and Circuit Level Considerations}}
The precision and reliability of analog computing operations depend heavily on individual memory cell properties. Key device parameters include:

\textbf{On/Off Ratio:} The ratio between maximum and minimum conductance/capacitance states defines the usable dynamic range for weight storage. Finite on/off ratios introduce non-zero current/charge from off-state cells that accumulates across array columns and can degrade accuracy, especially in larger arrays. To address this, designs often include ‘dummy’ columns (supported in NeuroSim \cite{peng_dnnneurosim_2019}) whose outputs are subtracted to cancel accumulated off-state current/charge. While effective, this slightly reduces computation dynamic range, resulting in smaller sensing margins and potentially smaller signal-to-noise ratio (SNR). Thus, high on/off ratios remain desirable for ACIM devices.

\textbf{Number of States:} Supporting high-precision weights requires either multi-level memory cells (MLC) or bit slicing with multiple lower-precision/binary cells. For MLC devices, the number of reliable programmable states determines weight precision. RRAM, PCM, Flash, and nvCap typically achieve 1-4 bits per cell \cite{cheng-xin_xue_241_2019, wan_compute--memory_2022, vinay_joshi_accurate_2020}, while FeFETs have demonstrated $>$5 bits \cite{jerry_ferroelectric_2017, aabrar_thousand_2022}.

At the circuit level, array size is limited mainly by sensing margins, as reliable output detection requires sufficient separation between analog states. This becomes challenging with larger arrays due to increased noise accumulation and reduced signal levels from parasitics.
Peripheral circuits include WL/BL drivers, ADCs, shift-and-add circuits, inter-array accumulation units, and multiplexers for data routing. Peripheral digital circuit blocks are verifiable via conventional VLSI methods; thus our behavioral modeling focuses primarily on analog computational elements.

\subsection{\textbf{Precision and Data Representation}}
Neural network parameters must be carefully mapped to CIM arrays, managing precision and computational accuracy. This presents two key challenges: representing multi-bit input activations and storing high-precision weights. For input activations, three main approaches exist:

\begin{enumerate}
    \item Multi-bit DACs convert inputs to analog voltages but require significant area/energy and may introduce device non-linearity errors.
    \item Bit-serial processing eliminates DAC overhead at the cost of multiple computational cycles.
    \item Time-encoding \cite{pwm_nand, timaq, yuyao} offers simplified ADCs at the cost of latency. Time-domain operation is not currently supported in NeuroSim.
\end{enumerate}

For weight storage, two methods are available:
\begin{enumerate}
    \item Bit Slicing: Used when desired weight precision ($b_w$) exceeds memory cell bit precision ($b_{cell}$). E.g., an 8-bit weight can be split across multiple memory cells.
    
    \item Direct Mapping: Used when weight precision ($b_w$) is less than or equal to cell precision ($b_{cell}$). In this case, $N_{cell}=1$, and the entire analog weight maps directly to a single memory cell's state.
\end{enumerate}

\begin{figure*}[t]
\centering
\includegraphics[width=\linewidth]{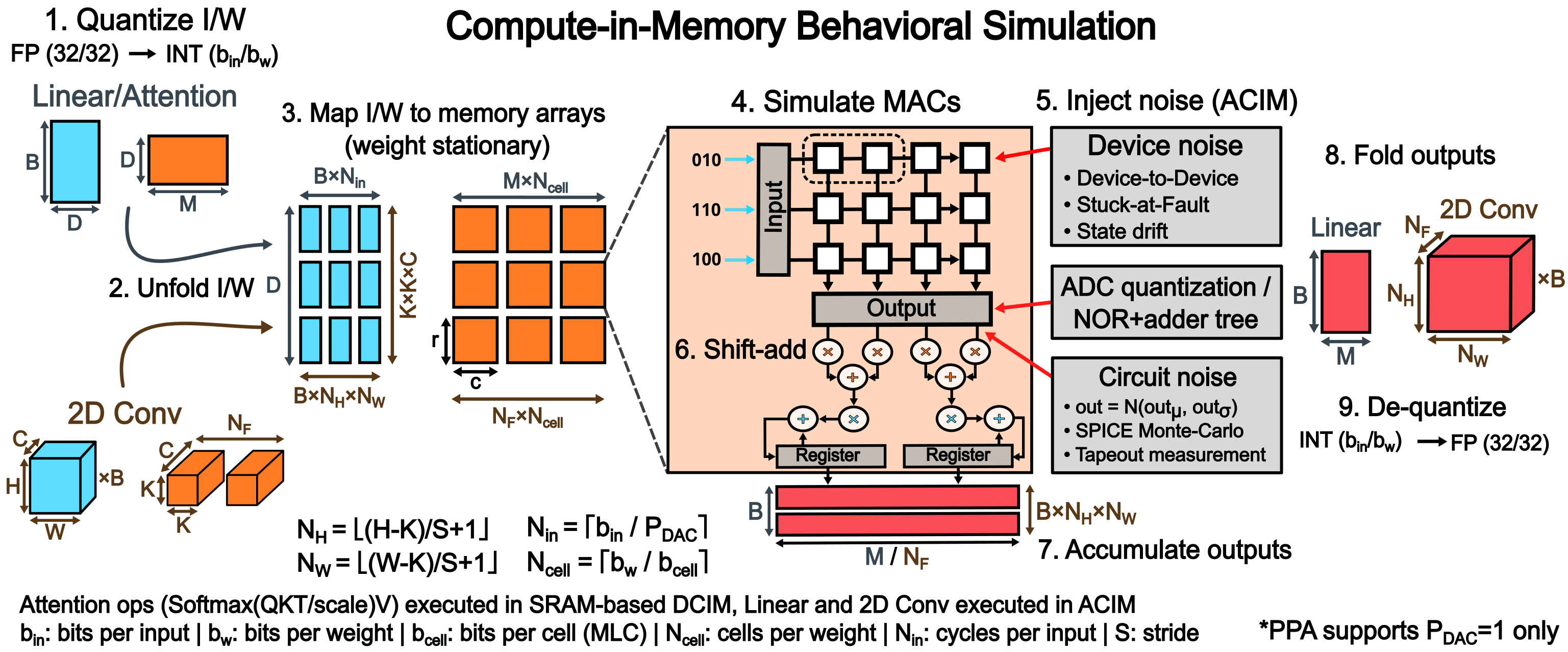}
\caption{Procedure for CIM behavioral simulation in NeuroSim V1.5. (1) Inputs and weights are quantized from floating point to integer. (2) Inputs and weights unfolded and (3) mapped to compute-in-memory arrays. (4) MACs are calculated with analytical circuit models. (5) Noise is injected either to memory devices (pre-ADC) or to digital MAC outputs (post-ADC). (6) ADC/Adder tree outputs are shifted and added. (7) Outputs from memory arrays are accumulated and concatenated. (8) Accumulated outputs are folded into output tensors. (9) Outputs are de-quantized from integer back to floating-point.}
\label{fig:hierarchy}
\end{figure*}

Bit-sliced weights occupy adjacent columns; outputs are scaled and summed post-ADC ('shift-add', Fig. \ref{fig:hierarchy}). The PPA estimator models bit-serial inputs for hardware accuracy. Pre-ADC analog shift-add for multi-bit weights \cite{jiang_enna_2023, guo_343_2024} and pre-ADC input accumulation schemes \cite{imagine, aacim} can reduce compute cycles due to bit-slicing. While we have explored these techniques \cite{hur_nonvolatile_2022}, they are not explicitly supported in this version but can be added with minimal user modification.

CIM MACs operate in integer domain, requiring float-to-int quantization. ADCs digitize analog results, which are then de-quantized to floating-point. Some accelerators explore quantizing the entire network to integers \cite{sehoon_kim_i-bert_2021, zhikai_li_i-vit_2022}, requiring specialized approximations for operations like normalization and activation functions.

\begin{figure*}[t]
\centering
\includegraphics[height=6cm,width=1\linewidth]{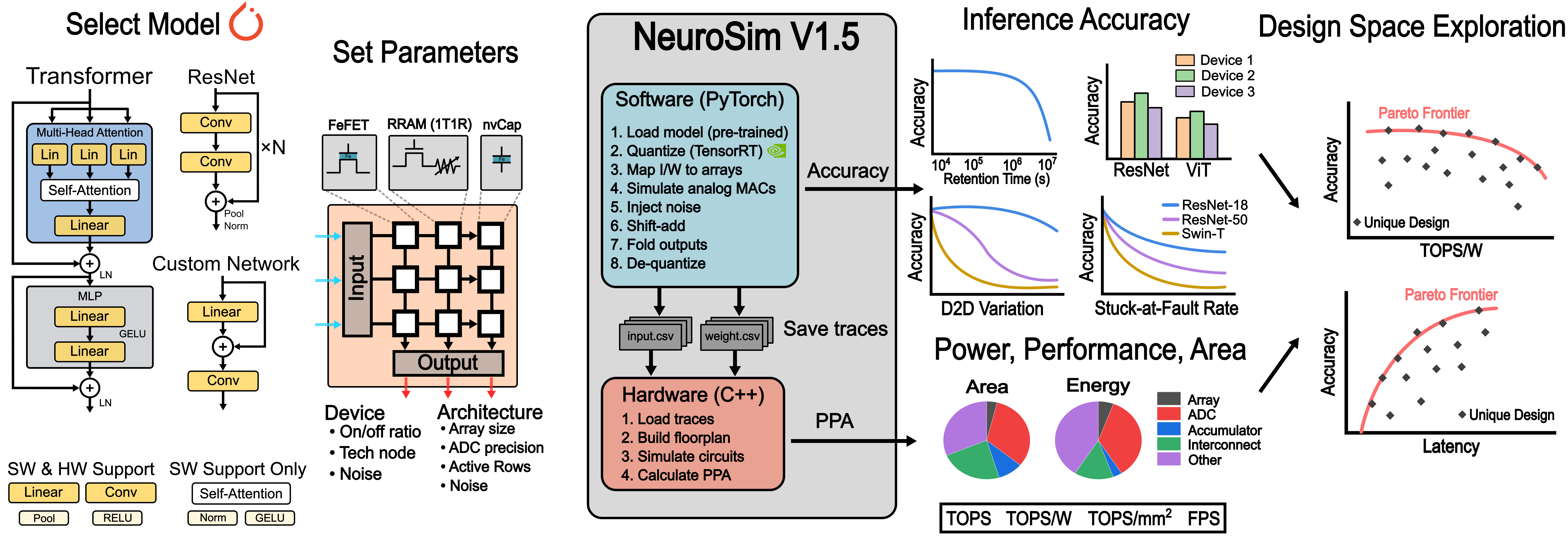}
\caption{Overview of NeuroSim V1.5: An integrated framework combining behavioral simulation with circuit-level PPA estimation for CIM accelerator design. NeuroSim V1.5 enables users to evaluate inference accuracy across diverse architectures, including CNNs and transformers, while simulating various device types and noise sources. This integration enables design-space exploration of the pareto frontier between accuracy and performance. Key updates include: hybrid ACIM/DCIM architecture support for transformers, flexible noise modeling via device expert and circuit expert modes, TensorRT-based quantization, expanded device support including nvCap, and up to 6.5$\times$ runtime improvement through GPU-accelerated simulation.}
\label{fig:motivation}
\end{figure*}

\subsection{\textbf{Hierarchical System Architecture}}
CIM accelerator organization stems from device technology and neural network structure constraints. Memory array sizes are typically $32\times32$ to $256\times256$ cells due to sensing margins and manufacturing yields. However, neural network layers often require much larger matrix operations, necessitating hierarchical decomposition.

NeuroSim's chip hierarchy: At base level, crossbar arrays compute partial sums of larger linear/convolutional layers. Crossbars are grouped into processing elements (PEs), where partial sums are accumulated. PEs are then grouped into tiles which include activation buffers and accumulation circuits for combining PE results-- enabling efficient large layer processing by distributing computations while maintaining data locality. At chip level, multiple tiles operate via an H-tree or X-Y bus interconnect. This highest level includes global buffers for inter-tile data movement and accumulation circuits for distributed computation results. The system implements layer-wise pipelining, where different tiles process consecutive network layers simultaneously to maximize throughput.

NeuroSim assumes all weights fit on-chip in NVM, enabling a fully weight-stationary dataflow. This design avoids frequent reprogramming that would stress NVM write endurance and increase latency. Small models that would underutilize a design's compute capacity are duplicated to boost throughput \cite{8702715}. For models exceeding single-die capacity, we have developed separate frameworks for 3D chiplet integration \cite{xiaochen_peng_heterogeneous_2021}.

The PPA estimator generates hardware configurations optimized to a specific network architecture (e.g. CNNs). While the floorplan is generated based on one selected network architecture, the hardware can execute multiple models of the same type through runtime reconfigurable weight mapping and dataflow control, as demonstrated in our prior work \cite{lu_runtime_2021}.

To simulate latency, area, and energy consumption, NeuroSim uses analytical models (C++) of circuit blocks based on standard cell libraries in advanced logic nodes. V1.4 \cite{lee_neurosim_2024} expanded technology support to 1nm, including stacked nanosheet transistors, as well as support for future A5 node with complementary-FETs. NeuroSim’s software/hardware co-design philosophy and open-source model enable research on emerging devices, array architectures, on-chip training \cite{peng_dnnneurosim_2021}, and advanced packaging \cite{xiaochen_peng_heterogeneous_2021, manley_cooptimization_2025, li_h3datten_2023}. For a more detailed hierarchy analysis, see \cite{peng_dnnneurosim_2019}.

\section{Simulation Methodology}\label{methodology}
\subsection{Overview of Design Methodology}
Designing CIM accelerators requires optimization across multiple abstraction layers—from device characteristics to system architectures. NeuroSim V1.5 provides an integrated framework to evaluate these designs through two main components (Fig. \ref{fig:motivation}) working jointly:
\begin{enumerate}
    \item A behavioral simulator that maps quantized neural networks to the CIM floorplan and evaluates inference accuracy under hardware non-idealities.
    \item A hardware analyzer that estimates system-level energy, latency, and area through detailed circuit models of arrays, peripherals, and interconnects.
\end{enumerate}

The behavioral simulator supports two primary simulation modes:
\textbf{Device expert mode:} Model memory array physics directly (conductance variations, stuck faults, temporal drift) to isolate memory technology effects on network accuracy.
\textbf{Circuit expert mode:} Model memory arrays statistically using pre-characterized noise distributions (SPICE/silicon), enabling rapid evaluation of aggregated device/circuit non-idealities.

Through TensorRT's post-training quantization, NeuroSim V1.5 automatically maps pre-trained neural networks to the corresponding CIM architecture. Subsequent sections detail quantization, array modeling, and incorporating non-idealities.

\subsection{Neural Network Quantization and Mapping}\label{quantization_and_mapping}
\subsubsection{\textbf{Quantization}}
NeuroSim V1.5 uses TensorRT's post-training quantization, unlike previous WAGE training-based methods \cite{wu_wage_2018}. TensorRT automatically replaces operators (linear, convolution, pooling) with configurable precision (e.g., 8b/8b, 6b/6b).

Scaling factors use max or histogram calibration (99.99\% CDF percentile, 2 batches).

\subsubsection{\textbf{Network to Array Mapping}}
Physical CIM array constraints require careful neural network operation partitioning. For example, for a linear layer (input $N$, output $M$), the quantized weight matrix maps to multiple arrays based on array dimensions and precision. The number of memory cells per weight depends on cell bit precision ($b_{cell}$). E.g., an 8-bit weight using 1-bit cells needs $N_{cell} = \lceil 8/1 \rceil = 8$ cells. Shown for a linear layer in Fig. \ref{fig:hierarchy}, these weight components split among adjacent columns, each column representing a different bit significance.

Memory array size (R rows, C columns) further constrains mapping. An $N \times M$ weight matrix partitions into $\lceil N/R \rceil$ array rows. Total arrays for weights: $\lceil N/R \rceil \cdot \lceil M N_{cell}/C \rceil$. To map 3D convolutional layers to 2D matrices we used im2col-like mapping described in \cite{peng_dnnneurosim_2019}.

The input feature map unfolds and maps per selected input scheme. Fig. \ref{fig:hierarchy} shows bit-serial operation (feature map processed separately per quantized input bit). Outputs from vertically-aligned arrays accumulate (partial sums for same output neurons). Outputs from horizontally-aligned arrays shift and add by bit significance, then concatenate for complete layer output.

\begin{figure}[t]
\centering
\includegraphics[width=0.95\linewidth]{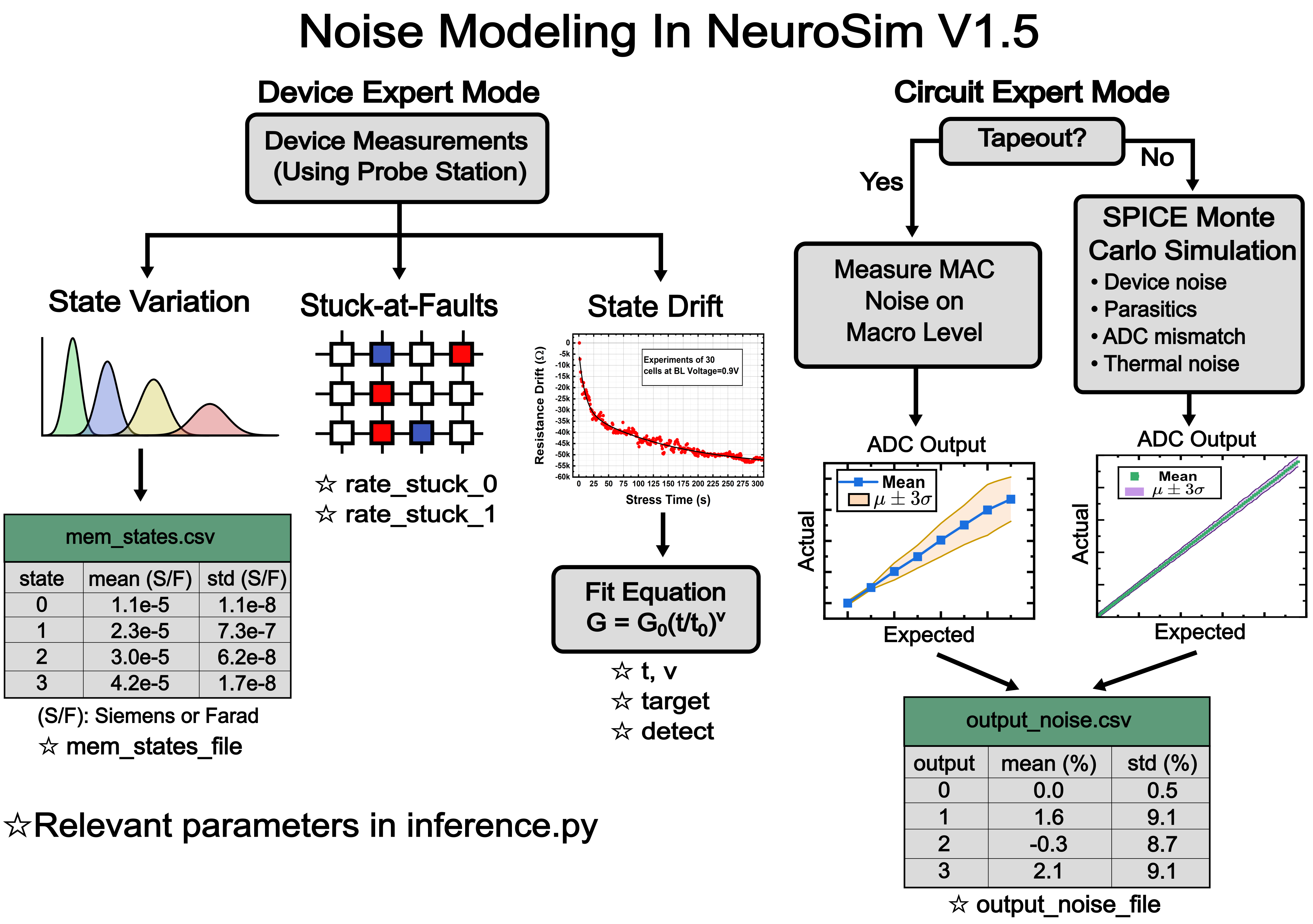}
\caption{NeuroSim V1.5 noise modeling flowchart. Device expert mode: user provides device data (memory state variation, stuck-at-faults, state drift). Circuit expert mode: user provides statistical data on MAC outputs from SPICE Monte-Carlo simulations or tapeout measurements.}
\label{fig:universal_modeling}
\end{figure}

\subsection{ACIM Behavioral Modeling}\label{behavioral_modeling}
\subsubsection{\textbf{Memory Array Operation}}
NeuroSim V1.5 replaces neural network layers with specialized CIM operators. Runtime improvements stem from: (1) optimized GPU-accelerated MVM simulation in `cim.Linear` and `cim.Conv` modules, and (2) efficient TensorRT quantization. Linear and convolutional layers auto-replace with custom modules that partition operations and simulate analog computation (Fig. \ref{fig:universal_modeling}).

Two simulation modes exist: device expert (measured variations, aging models) and circuit expert (SPICE/silicon statistics), used separately to avoid double-counting.

\subsubsection{\textbf{Device and Circuit Non-Idealities}}
In device expert mode, integer weights map to conductance/capacitance states. The framework simulates analog multiplication via Ohm's law ($I=GV$) for resistive arrays or charge-based computation ($Q=CV$) for capacitive arrays. Inputs are idealized voltages; outputs are summed column currents/charges. Noise reflecting device non-idealities is typically injected at this stage, pre-ADC quantization.

To capture all possible input and weight bit-slicing combinations, including both bit-serial and DAC-based input schemes, we use the following generalized equation:

\begin{equation}
\label{eq:critical_loop}
y = \sum_{i}^{N_{cell}} \sum_{j}^{N_{in}} (2^{i\cdot b_{cell}}) (2^{j \cdot P_{DAC}}) (W_i\cdot x_j)
\end{equation}

where:

\begin{itemize}
    \item $b_{in}$ is the bit significance of the inputs.
    \item $P_{DAC}$ is DAC precision ($P_{DAC}=1$ for bit-serial).
    \item $N_{in}$ is the number of input cycles $\lceil b_{in} /P_{DAC}\rceil$
    \item $b_{w}$ is the bit significance of the weights.
    \item $b_{cell}$ is the number of bits per weight cell (e.g. $b_w=4$ for 4-bit MLC)
    \item $N_{cell}$ is the number of cells per weight $\lceil b_w / b_{cell} \rceil$
    \item $W_i$ represents the $i$-th bit slice of the weight matrix
    \item $x_j$ represents the $j$-th bit slice of the input vector
\end{itemize}

For example, when using 4-bit MLC ($b_{cell}=4$) and 8-bit weights ($b_w=8$), $N_{cell}=2$ so $i$ ranges from 0 to 1 and the weight scales for each iteration are $2^{i\cdot b_{cell}} = $ $2^{0 \cdot 4} = 1$ and $2^{i\cdot b_{cell}} = $ $2^{1 \cdot 4} = 16$. When using bit-serial mode ($P_{DAC}=1$) to represent 8-bit activations ($b_{in}=8$), $N_{in}=8$ so $j$ ranges from 0 to 7 with input scales increasing by powers of 2. Each sum in equation \ref{eq:critical_loop} corresponds to a for-loop in the behavioral simulation. Consequently, the nested loop becomes the critical piece of code to optimize. Using PyTorch's GPU-accelerated tensor operations, we can calculate the output $y$ of every memory array on the CIM chip design in parallel, which greatly reduces the overall runtime of NeuroSim.

Shown in Fig. \ref{fig:universal_modeling}, in device expert mode, three primary categories of device variations are modeled:

\textbf{Device-to-Device (D2D) Variation}: For each conductance (or capacitance) state $i$, the programmed value follows a Gaussian distribution:
\begin{equation}
    G = N(G_{mean_i}, \sigma_i)
\end{equation}

where $G_{mean_i}$ is the target conductance for state $i$ and $\sigma_i$ is the user-specified standard deviation. Separate variation parameters are specified for each level $i$. Users provide these parameters as comma-separated tuples, with one variation value per memory state. Users store these in the file ‘mem\_states.csv’ that is then read by the framework automatically.

\textbf{Stuck-at-Faults (SAF)}: Cells permanently fixed at min/max states due to defects. Modeled by randomly setting cells per user-specified probabilities during initialization, unlike temporal variations.

\textbf{Temporal Drift}: Time-dependent conductance changes follow:
\begin{equation}
\label{eq:drift}
    G(t) = G_0(t/t_0)^v
\end{equation}

where $G_0$ is initial conductance, $t$ is retention time, $v$ is the drift coefficient, and $t_0$ is a reference time. The framework supports two drift modes: random drifting, or drifting towards a fixed state. The simulation treats arrays as frozen at the specified time point $t$ for inference calculations.

Circuit expert mode uses statistical models from Monte Carlo SPICE (varying device properties, temperature, voltage) or silicon measurements (repeated column measurements). Both produce mean/standard deviation pairs for each MAC output, visualized as confusion matrices (Fig. \ref{fig:universal_modeling}) and stored in 'output\_noise.csv'.

This approach skips the critical loop (Eq. \ref{eq:critical_loop}), calculating ideal partial sums with minimal overhead, then sampling MAC distributions for noise injection. Enables rapid evaluation with empirically validated characteristics (Section \ref{case_studies}).

While this statistical approach captures the aggregate impact of these effects, it doesn't model detailed physics per input/weight. This results in a trade-off in physics-accurate simulations for fast large-scale design exploration. For highly detailed analysis of such effects, tools like CrossSim \cite{xiao_tianyao_crosssim_2021} may be more suitable. 

Additionally, while the current implementation samples device variations with spatial-independence, users can model spatially-correlated defects by supplying the desired variations for those areas. This would require supplying the variation data explicitly for the memory cells in the area of interest along with slight modification to the error injection. Such a feature would be useful for post-silicon measured defects and may be explicitly supported in the future.

\subsection{Performance Analysis Framework}
NeuroSim V1.5 uses a trace-based interface for PPA estimation. The behavioral simulator saves quantized inputs/weights as CSV traces (Fig. \ref{fig:motivation}) capturing hardware data patterns.

The C++ PPA estimator auto-generates accelerator floorplans and creates transistor-level circuit models (arrays, ADCs, interconnects, peripherals) calibrated with SPICE and tapeout data \cite{9755965}. Using traces, it calculates per-array energy with explicit bit-serial modeling. Weights convert to conductance/capacitance states per configuration parameters.

In NeuroSim V1.5, we extended the hardware analyzer to support charge-based computation with non-volatile capacitive memories. This required the addition of new circuit models for nvCap arrays, charge-to-voltage conversion using an amplifier \cite{luo_design_2021}, and control circuits for the unique charge-based computation. 

\subsection{\textbf{Support for Vision Transformers}}
\begin{figure}[t]
\centering
\includegraphics[width=0.95\linewidth]{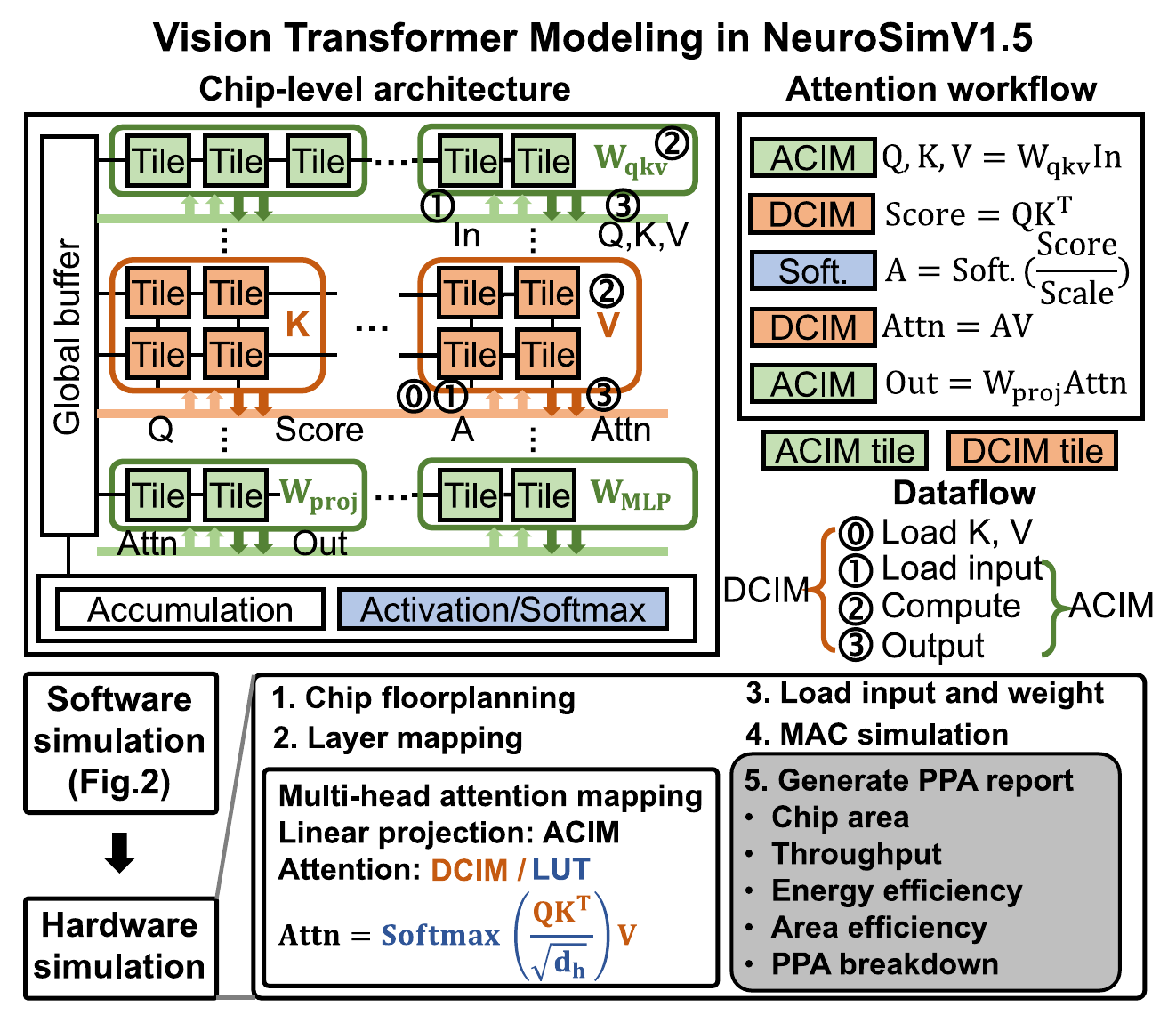}
\caption{NeuroSim V1.5 ViT dataflow. Linear projections execute on ACIM tiles; attention score and aggregation use DCIM tiles. Numbered circles indicate the operation sequence within one transformer block. Layer-level pipelining maps different layers to different tiles, enabling all tiles to execute in parallel on different images.}
\label{vitModeling}
\end{figure}

Vision transformer (ViT) architectures are supported in NeuroSim V1.5 using a hybrid ACIM/DCIM approach described in Fig. \ref{vitModeling}. Linear layers (Q, K, V projections, MLPs) use ACIM arrays in weight-stationary mode. Attention score computation ($QK^T$) and weighted aggregation use DCIM arrays, as these operations require writing intermediate attention scores as operands for subsequent multiplication---incompatible with NVM write latency and endurance constraints.

The hybrid architecture maintains NeuroSim's existing tiled hierarchy. Entire tiles are dedicated to either ACIM or DCIM, with identical bit-serial input processing. Multi-head attention maps naturally to this structure: each head has independent projection weights, enabling parallel execution across tiles. Different layers map to different tiles, and layer-level pipelining allows all tiles to operate simultaneously on different images, maximizing hardware utilization. Outputs from each tile are written to the global buffer before being sent to the next layer. The global buffer is sized to fit the appropriate activation data from each tile operating in parallel. Softmax and GELU are implemented via 8-bit lookup tables (LUTs), storing input-output relationships for the quantized operators.

The DCIM array architecture follows \cite{chih_164_2021} with circuit modeling from \cite{lee_neurosim_2024}. NeuroSim V1.5 supports both noise-integrated accuracy simulation and PPA evaluation for ViT models.

\begin{table}[htbp]
\centering
\begin{threeparttable}
\caption{Design Space Exploration Parameters and Device Non-idealities Case Study Parameters}
\label{design_space_params}{
\renewcommand{\arraystretch}{1.4}
\begin{tabularx}{8.5cm}{>{\centering\arraybackslash}m{2.5cm} >{\centering\arraybackslash}m{2.5cm} >{\centering\arraybackslash}m{2.5cm}}
\toprule
\textbf{Parameter}         & \textbf{Range / Values for DSE} & \textbf{Values for Noise Case Studies} \\ 
\midrule
Technology Node           & \multicolumn{2}{c}{22 nm}\\
Device                    & \multicolumn{2}{c}{RRAM}\\
HRS / LRS*                & \multicolumn{2}{c}{40k$\Omega$ / 3k$\Omega$}\\
Input Processing           & \multicolumn{2}{c}{bit-serial} \\
Memory Cell Precision      & \multicolumn{2}{c}{1b $-$ 4b} \\
Network / Dataset & ResNet-18 / CIFAR-100 & \makecell{\parbox[c]{2.5cm}{\centering VGG8 / CIFAR-10 \\ ResNet-18 /\\CIFAR-100 \\ ResNet-50 /\\ImageNet 1k \\ Swin-T /\\ImageNet 1k}}\\
Array Dimensions           & $32\times 32$ - $256\times 256$ & $128\times 128$\\ 
Rows Active                & all & 8 / 32 / 128 \\
Input / Weight Precision   & 6b, 8b & 8b  \\  
ADC Precision**              & 3b $-$ 11b & full precision   \\ 

\bottomrule
\end{tabularx}
}

\begin{tablenotes}
    \item * RRAM parameters based on Intel's 22 nm FinFET platform \cite{jain_132_2019}
    \item ** ADC full precision is calculated using equations (\ref{out_max}) and (\ref{PADC})
\end{tablenotes}
\end{threeparttable}
\end{table}





\subsection{Evaluation Methodology}
Our experimental evaluation consists of three main components: design space exploration of hardware configurations, inference accuracy analysis under various device non-idealities, and runtime performance comparison across open-source platforms. The DNN algorithms and proposed NeuroSim framework are implemented using PyTorch, with all simulations performed on a single NVIDIA RTX A6000 GPU with 48GB GDDR6 memory.

\subsubsection{\textbf{Design Space Exploration}}
For the design space exploration, we examine the trade-offs between hardware efficiency and inference accuracy across a range of architectural parameters. Table \ref{design_space_params} summarizes the key parameters and their ranges used in the design space exploration study. The memory device characteristics are based on Intel's 22 nm RRAM platform \cite{jain_132_2019}. We process inputs using the bit-serial approach (1b per cycle) to maintain consistency with the hardware estimator and activate all rows in parallel to maximize throughput.

The ADC precision requirements are determined by the dynamic range of array outputs, which depends on three key factors: memory cell precision ($b_{cell}$), DAC precision ($P_{DAC}$), and array rows ($R$). For a given array configuration, the maximum output value can be calculated as:

\begin{equation}
\label{out_max}
    out_{max} = R(2^{P_{DAC}}-1)(2^{b_{cell}}-1)
\end{equation}

The minimum required ADC precision ($P_{ADC}$) to capture this full range without quantization loss is:

\begin{equation}
\label{PADC}
    P_{ADC}=\lceil log_{2}(out_{max}) \rceil
\end{equation}

In our design space exploration, we evaluate two ADC precision settings for each configuration:
\begin{enumerate}
    \item Full precision matching the calculated dynamic range ($P_{ADC}$).
    \item Reduced precision ($P_{ADC}-1$ and $P_{ADC}-2$) to study the impact of quantization noise.
\end{enumerate}

Excluding device and circuit noise, the inference accuracy is solely affected by the input/weight quantization from floating-point to fixed-point representation and the quantization of array outputs depending on the ADC precision. Design parameters were swept using automated scripts to explore the trade-offs in accuracy and performance considering input/weight/ADC quantization and array size. Results are shown in Fig. \ref{fig:design_space} and discussed  further in section \ref{results}.

When reducing ADC precision below the ideal precision, we adopt a clipping approach where the sensing margins for each analog output state remain the same regardless of ADC precision. Any outputs larger than the precision of the ADC are clipped to the maximum ADC output based on its precision. We find that this method results in comparable accuracy to dynamic quantization methods while also being the most practical to implement in hardware.

\subsubsection{\textbf{Device Noise Case Studies}}
For analyzing inference accuracy under device non-idealities, we standardize the hardware configuration to isolate the effects of various noise sources and variations. Table \ref{design_space_params} presents the fixed parameters used across all device non-ideality studies. 

In these case studies, we analyze different sources of device variation including device-to-device variation, stuck-at-faults, and state drift across multiple DNNs and datasets to demonstrate support for multiple networks and noise types. 

\subsubsection{\textbf{Circuit Noise Case Studies}}
Next, we demonstrate the power of our improved noise modeling by using circuit noise data characterized both from SPICE characterization and tapeout measurements. With this improved modeling, direct comparisons between different devices, compute circuits, and DNNs can be performed. The table in Fig. \ref{outputNoise} summarizes the key characteristics of each platform, with parameters drawn directly from their respective publications.

Mean and standard deviation for each output state are supplied in a comma separated format (.csv file), where each row of the file corresponds to the mean for each given output (post-ADC) and the standard deviation of the output.

\subsubsection{\textbf{Runtime Evaluation}}
To characterize the improvements in runtime over previous versions of NeuroSim, we record the runtime of the inference accuracy portion under several key conditions. These include the network size, input resolution, the MLC level, precision of the DAC (if one is used), and the input and weight precision.

\begin{table}[htbp]
\centering
\begin{threeparttable}
\caption{Default 22 nm RRAM PPA (ResNet-18 / CIFAR100)}
\label{tab:ppa}{
\renewcommand{\arraystretch}{1.4}
\begin{tabularx}{8cm}{
>{\centering\arraybackslash}X 
>{\centering\arraybackslash}X
>{\centering\arraybackslash}X
>{\centering\arraybackslash}X
>{\centering\arraybackslash}X}
\toprule
\textbf{TOPS} & \textbf{TOPS/mm$^2$}  & \textbf{TOPS/W} & \textbf{FPS} & \textbf{Accuracy} \\ 
\midrule
11.6             & 0.013 & 21.3 & 7770 & 75.57\%       \\ 
\bottomrule
\end{tabularx}
}

\begin{tablenotes}
    \item Example performance using 22 nm RRAM \cite{jain_132_2019} and default settings in NeuroSim: $128\times128$ array size, 7-bit ADC, 8b/8b input/weight precision.
\end{tablenotes}
\end{threeparttable}
\end{table}

\section{Results and Case Studies}\label{results}
\subsection{Design Space Exploration}

We demonstrate NeuroSim V1.5's design space exploration capability using a 22 nm RRAM \cite{jain_132_2019} platform. Following Table \ref{design_space_params}, we evaluate accuracy-efficiency tradeoffs across architectural choices including array size, precision, and data representation.

\begin{figure}[tbp]
   \centering
   \includegraphics[width=\linewidth]{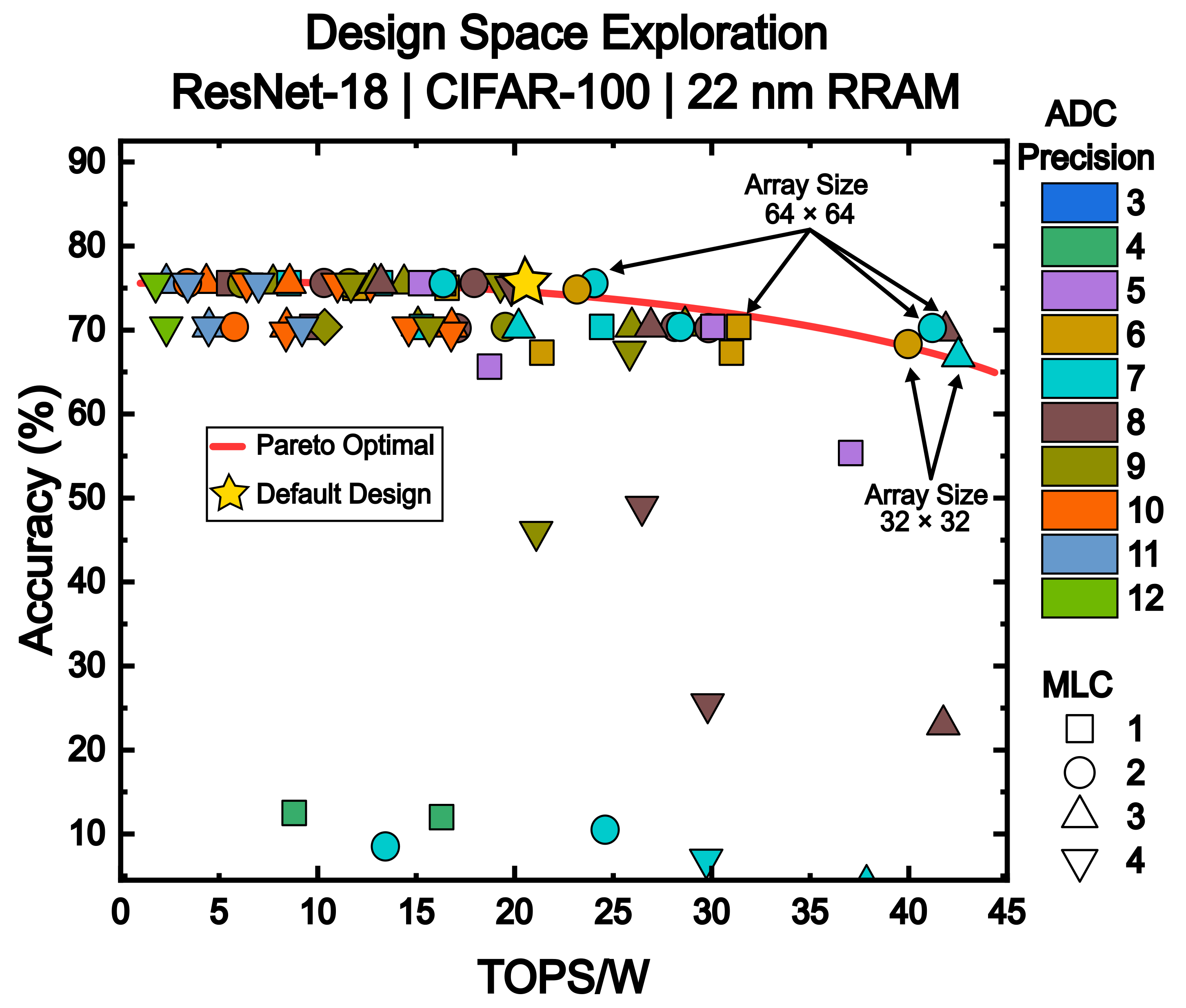}
   \caption{Example design space exploration for 22 nm RRAM. Each point represents a unique configuration. Bit-serial input was used for every design. Designs with highest TOPS/W use array sizes of $32\times 32$ and $64\times 64$.}
\label{fig:design_space}
\end{figure}

Table \ref{tab:ppa} shows example performance metrics for the default configuration: 11.6 TOPS at 21.3 TOPS/W while processing 7,770 frames per second for CIFAR-100 images. Systematic exploration (Fig. \ref{fig:design_space}) reveals designs that significantly outperform this baseline, identifying key trends for optimal CIM architectures:

\begin{enumerate}
    \item ADC Precision: Pareto-optimal designs cluster in 5$-$8 bit range. ADC precision can be reduced by 1-bit from lossless precision without significant accuracy loss, and by 2-bits in certain configurations.
    \item Array Dimensions: Optimal designs use array sizes between $32\times 32$ and $128\times 128$ cells. Smaller arrays reduce ADC precision requirements but need more accumulation circuitry, while larger arrays require larger ADCs, incurring latency and energy costs. Designs achieving $>$40 TOPS/W used $32\times 32$ or $64\times 64$ arrays.
    \item Memory Cell Precision: 2-bit and 3-bit MLC configurations dominate the Pareto frontier, balancing storage density and programming reliability while achieving highest energy-efficiency for any given accuracy target.
\end{enumerate}

Across the explored configurations, Pareto-optimal designs converge toward 6-7 bit ADC precision regardless of other parameter choices. This range represents the balance point between ADC overhead (energy, latency, area) and throughput. Multiple parameter combinations achieve similar efficiency at this ADC precision: 1-bit cells with 128 rows, 2-bit cells with 64 rows, or 4-bit cells with 32 rows. The choice between these configurations then shifts to device multi-state capability and rate of variations. Current ADC designs dominate the energy and latency budget in ACIM arrays. If more efficient ADC architectures become available—through techniques such as time-domain sensing, or quantization schemes for reduced-precision —the optimal design point would shift toward larger array sizes or higher-precision MLC devices that are currently constrained by ADC overhead. Next, we demonstrate the improved noise-modeling capabilities in the new behavioral simulator.

\begin{figure} [ht]
\centering
\includegraphics[scale=0.44]{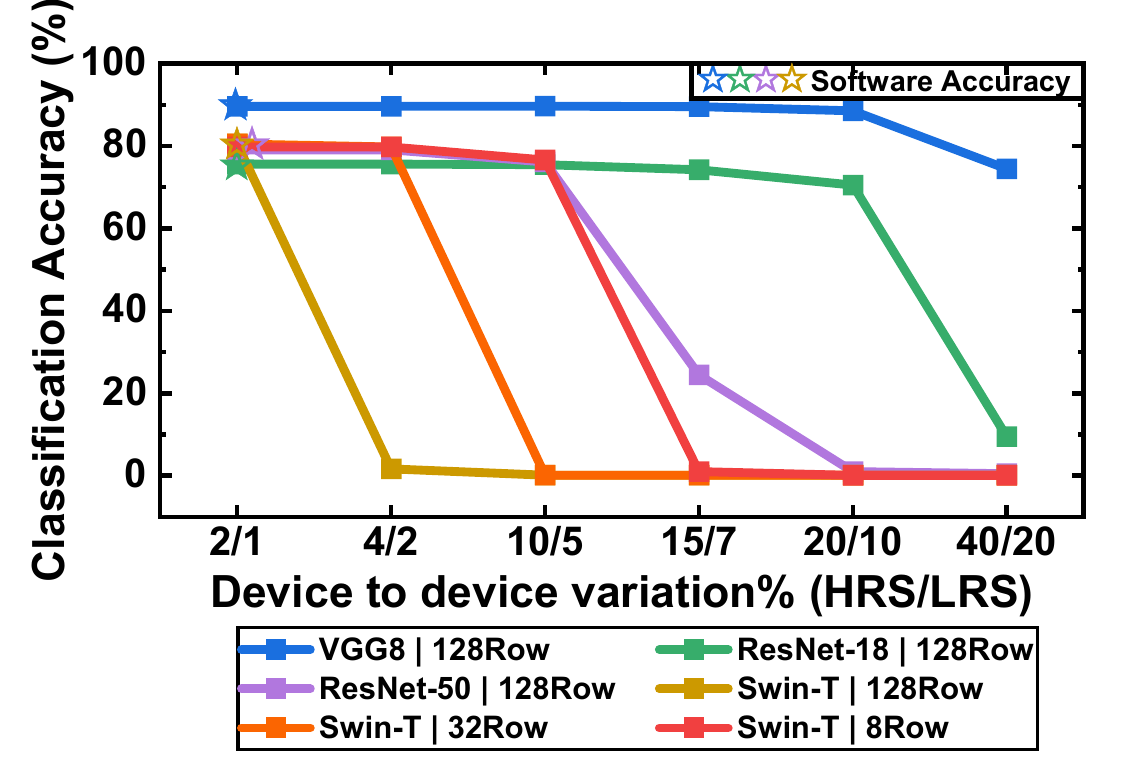}
\vspace{-3mm}
\caption{Inference accuracy of VGG8 on CIFAR-10, ResNet-18 on CIFAR-100, ResNet-50, and Swin-T on ImageNet with different degrees of 1b RRAM device variation. "XRow" describes the number of rows activated in parallel.}
\label{d2dvariation}
\end{figure}

\subsection{Noise Modeling Case Studies}\label{case_studies}
We evaluated inference accuracy across multiple neural networks using NeuroSim V1.5's noise modeling capabilities, including D2D variation, memory state drifting, stuck-at-faults and circuit-level MAC output noise. We selected network architectures of increasing complexity: VGG8 for CIFAR-10, ResNet-18 for CIFAR-100, ResNet-50 for ImageNet, and Swin Transformer Tiny for ImageNet. Following established methodology \cite{zhou2016dorefa}, we excluded the first convolutional layer from quantization.

\subsubsection{\textbf{D2D Variation}}
To explore the effect of memory state variation on inference accuracy, we use a 1-bit RRAM baseline design and select several variation rates in the HRS and LRS ranging from 1\% to 40\%. Based on previous measurement results, we use a higher variation in the HRS than the LRS.

Fig. \ref{d2dvariation} shows sensitivity is related with network complexity and dataset difficulty. VGG8 on CIFAR-10 maintains reasonable accuracy until LRS variation reaches 20\%, while ResNet-18 on CIFAR-100 and ResNet-50 on ImageNet show significant drops at just 10\% and 5\% LRS variation, respectively. Swin-T shows the lowest tolerance to D2D noise, dropping to 0\% accuracy at 4\%/2\% when using the same array and ADC precision as ResNet-50. To improve the accuracy of Swin-T, we decrease the array size to 32$\times$128 array configuration (vs. 128$\times$128 for CNNs), improving signal-to-noise ratio by reducing the number of simultaneously activated rows. This configuration also reduces DCIM adder tree overhead while maintaining comparable throughput, as shown in Table~\ref{tab:parallel_read}. We analyze the increased sensitivity of Swin-T in section \ref{vitErrorAnalysis}. 

\begin{figure} [ht]
\centering
\includegraphics[scale=0.44]{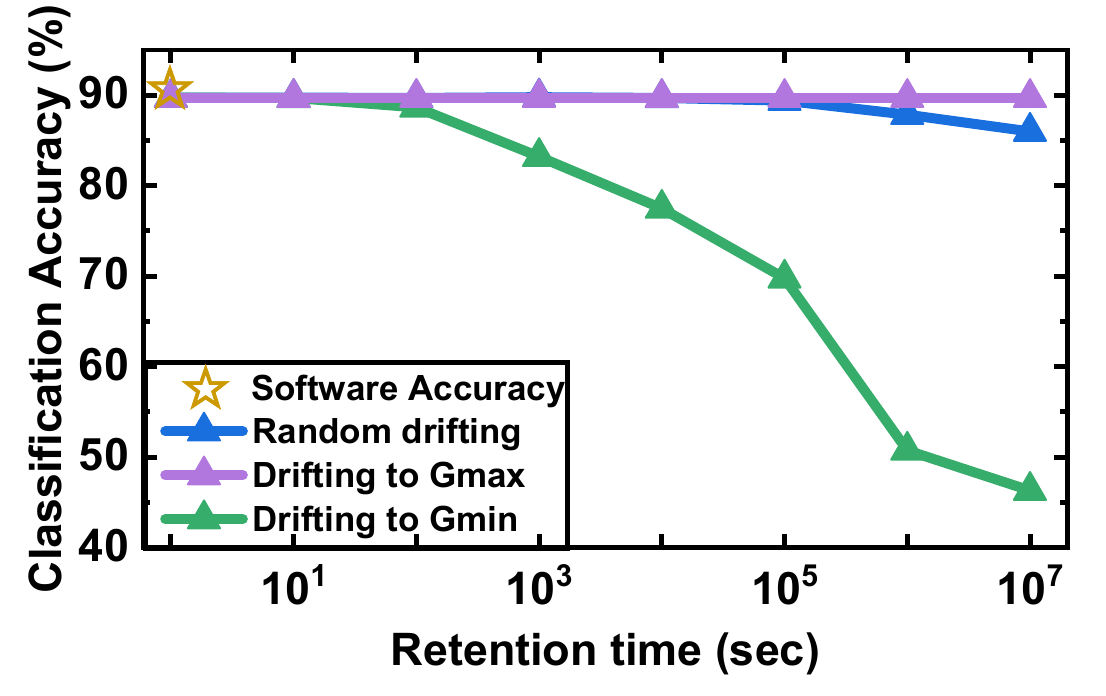}
\vspace{-4mm}
\caption{Inference accuracy of VGG8/CIFAR-10 regarding random device conductance drifting, drifting to Gmax, and drifting to Gmin with 1b cells.}
\vspace{-0.5mm}
\label{drifting}
\end{figure}

\subsubsection{\textbf{Conductance Drift}}
Fig. \ref{drifting} examines how three different types of conductance drift (modeled by Equation \ref{eq:drift}) affect inference accuracy. Drifting toward maximum conductance (Gmax) maintains the best accuracy retention. Random drifting shows the next best resilience; this occurs because cells at minimum conductance cannot drift lower, while cells at maximum conductance cannot drift higher, effectively limiting the total conductance change. Drifting toward minimum conductance (Gmin) results in the largest accuracy degradation. This asymmetry between Gmax and Gmin drift arises from Equation \ref{eq:drift}; for the same retention time $t$ and drifting speed $|v|$, drifting toward Gmin can result in a larger absolute conductance change (if starting from a higher state) compared to drifting toward Gmax (if starting from a lower state), leading to a larger deviation of MAC outputs.

\subsubsection{\textbf{Stuck-at-Faults}}
\begin{figure}[ht]
    \centering
    \includegraphics[scale=0.35]{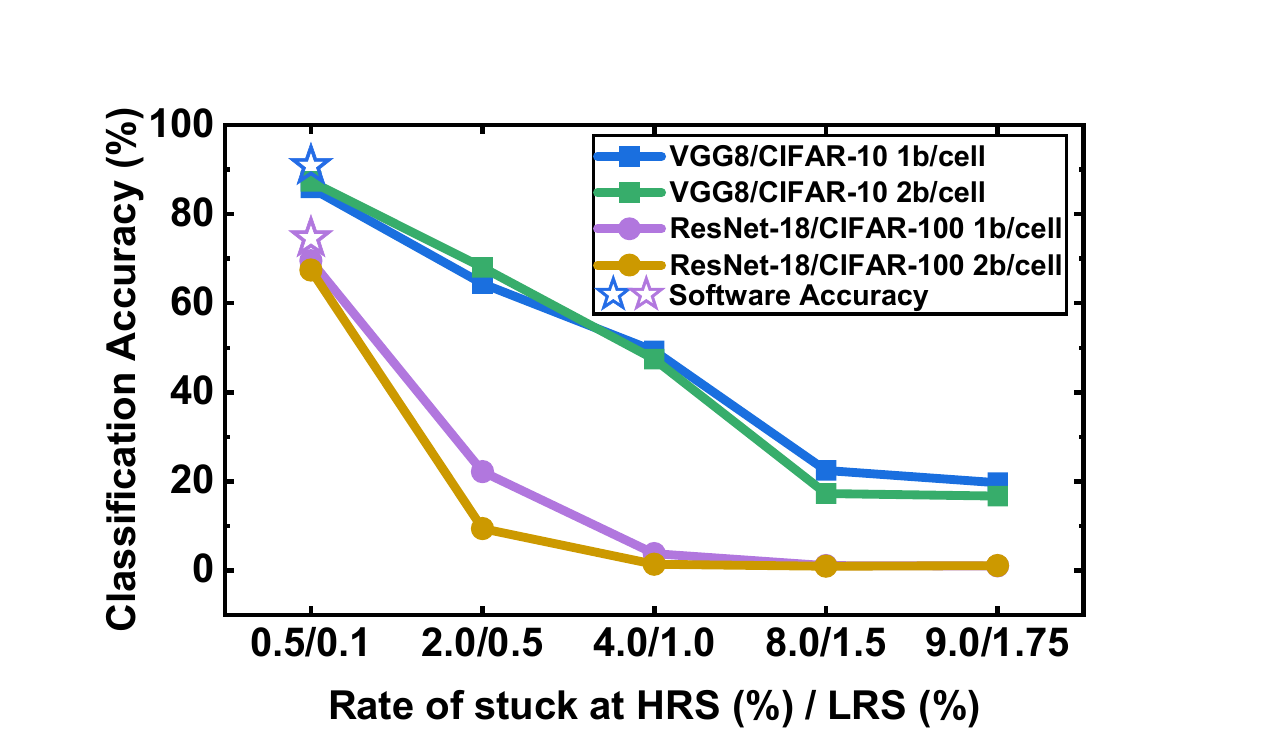}
    \vspace{-3mm}
    \caption{Inference accuracy of VGG8 and ResNet-18 under different stuck-at-fault rates (1b RRAM, 128x128 array, 7b ADC).}
    \vspace{-3mm}
    \label{stuckAtFault}
    \end{figure}

NeuroSim V1.5 adds stuck-at-fault (SAF) error modeling, extending beyond V1.4's D2D variation and drift modeling. Fig. \ref{stuckAtFault} shows SAF error impact on inference accuracy across different networks, with SAF rates bounded by realistic RRAM fabrication constraints: 1.75\% for LRS and 9.0\% for HRS \cite{he2019noise}.
Both 1-bit and 2-bit cells show similar accuracy degradation at equivalent SAF rates, suggesting cell precision does not significantly influence SAF sensitivity. However, network complexity strongly affects SAF tolerance - ResNet-50 on ImageNet loses accuracy even at minimal SAF rates, while VGG8 and ResNet-18 maintain limited functionality only at the lowest SAF rates.
SAF errors cause more severe accuracy degradation than other noise sources like D2D variation or conductance drift. This aligns with previous findings \cite{he2019noise}, highlighting the critical importance of minimizing stuck faults during fabrication.

\subsubsection{\textbf{MAC Output Noise}}
circuit-level noise sources during array computation can be captured as statistical variations in the post-ADC MAC outputs. To demonstrate NeuroSim V1.5's flexible noise modeling capabilities, we evaluated inference accuracy across four different CIM macros, incorporating both SPICE-characterized circuit data and silicon measurements from fabricated chips.

CIM A and B \cite{yixin_multilevel_2024} employ 2b FeFET technology with different computing modes: CIM A performs current-mode computation with FeFETs, while CIM B implements charge-based computation using FeFET-modulated capacitors. Their noise characteristics were derived from Monte-Carlo SPICE simulations. CIM C \cite{9755965} uses silicon measurements from a fabricated RRAM-based CIM chip. For these three designs, each MAC output level is characterized by unique mean and standard deviation values. CIM D \cite{luo_cross-layer_2024, 10237236} uses 28 nm nvCap with charge-mode computation, where SPICE simulations revealed thermal noise as the dominant source, here, a uniform variation across output levels was used.

\begin{figure}[htb]
\centering
\includegraphics[scale=0.33]{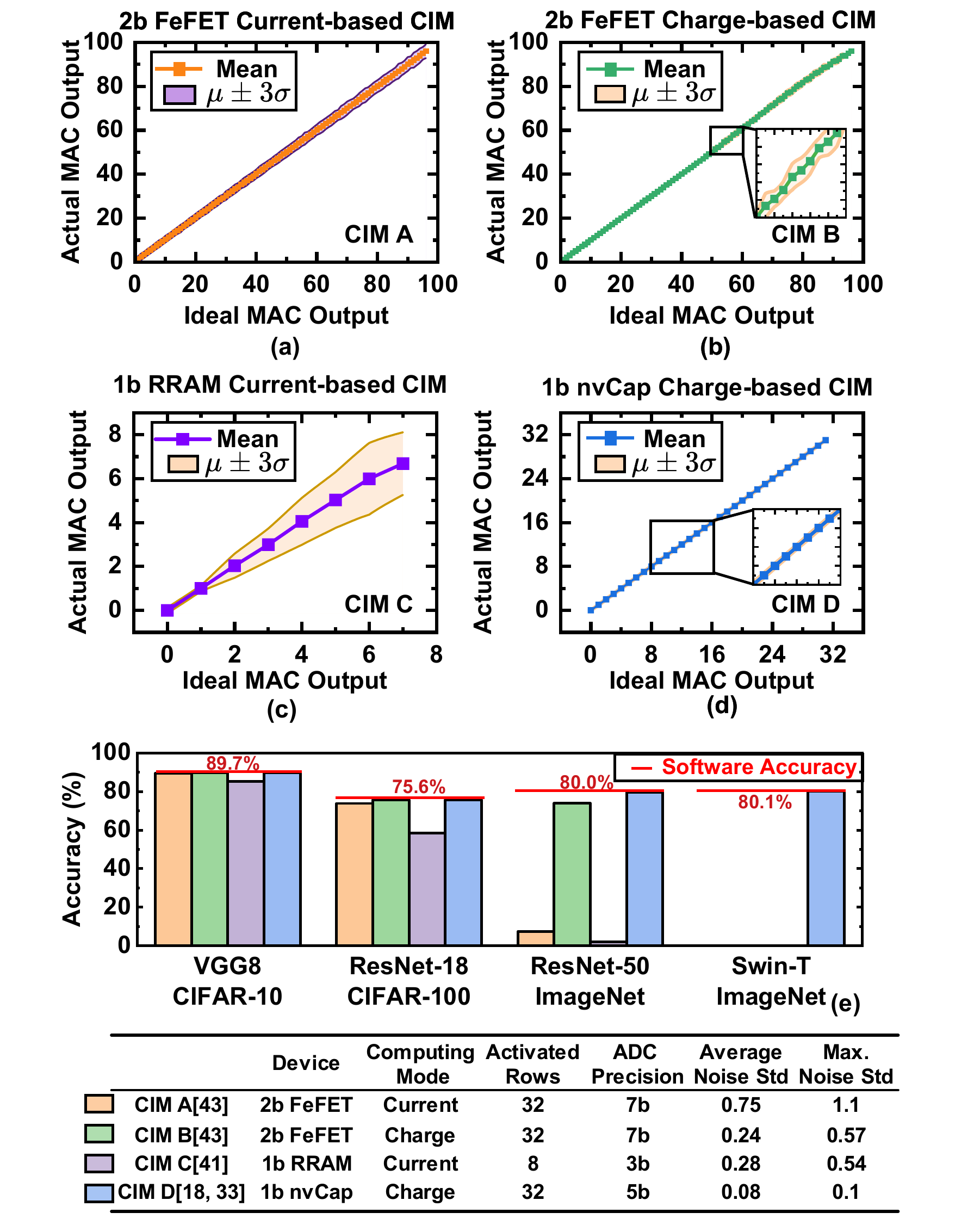}
\vspace{-2mm}
\caption{(a) Actual MAC output with 2b FeFET current domain CIM \cite{yixin_multilevel_2024} under 50mV Vth variation; (b) Actual MAC output with 2b FeFET charge domain CIM \cite{yixin_multilevel_2024} under 50mV Vth variation; (c) Actual MAC output of an RRAM tape-out \cite{read_enabling_2023}; (d) Actual MAC output with 1b nvCap charge domain CIM \cite{luo_cross-layer_2024, 10237236} under 50mV Vth variation; (e) Inference accuracy of various neural network algorithms across CIM macro A-D.}
\label{outputNoise}
\end{figure}

Figs. \ref{outputNoise}(a-d) illustrate the statistical distribution of actual versus ideal MAC outputs for representative cases. While CIM A, B, and D show tighter error bounds compared to CIM C, all designs achieve similar accuracy on smaller networks. Fig. \ref{outputNoise}(e) compares inference accuracy across architectures of increasing complexity. VGG8 and ResNet-18 maintain reasonable accuracy across all implementations, with CIM C showing slightly higher degradation (\>2\%) due to increased output variation. For ResNet-50, only CIM B and D - which achieve lower standard deviations in their MAC errors - maintain high accuracy. The Swin Transformer exhibits particular sensitivity to output noise, with only the nvCap-based CIM D achieving software-comparable performance.

These results also validate the circuit expert mode's statistical modeling approach. By characterizing MAC output distributions from SPICE simulations or silicon measurements and injecting noise accordingly, the framework captures the aggregate impact of device and circuit non-idealities without requiring per-cell physical simulation. By providing a standardized methodology for capturing MAC output statistics, the framework enables direct comparison of different technologies and compute circuits under realistic operating conditions.

\subsection{Noise Sensitivity Analysis of ViTs}\label{vitErrorAnalysis}

As shown in Fig. \ref{d2dvariation}, Swin-T exhibits lower tolerance to analog noise compared to ResNet-50. To understand the underlying mechanisms, we analyze errors at both the layer and circuit levels using device-to-device (D2D) variations as the primary noise source.

\subsubsection{\textbf{Related Work on Transformer Sensitivity}}
Prior work attributes transformer sensitivity in analog CIM to non-uniform activation distributions causing quantization challenges \cite{zhikai_li_i-vit_2022, sehoon_kim_i-bert_2021}, heightened input/output interface noise susceptibility \cite{gallo_using_2023}, and MSB-cycle error amplification in bit-serial computation \cite{zhang_asim_2025}. 

\subsubsection{\textbf{Layer-wise Error Analysis}}
To identify whether specific layers or architectural components drive Swin-T's higher noise sensitivity, we compute the root mean square error (RMSE) in each layer output under the 4\%/2\% D2D variation condition. RMSE is computed as $\sqrt{\text{mean}((\hat{y} - y)^2)} / \sqrt{\text{mean}(y^2)}$, where $y$ is the ideal floating-point layer output and $\hat{y}$ is the noisy output. Tests were conducted for both ResNet-50 and Swin-T on ImageNet with 1-bit cells, an array size of 128$\times$128, and 7-bit ADC. 

\begin{figure}[t]
\centering
\includegraphics[width=\linewidth]{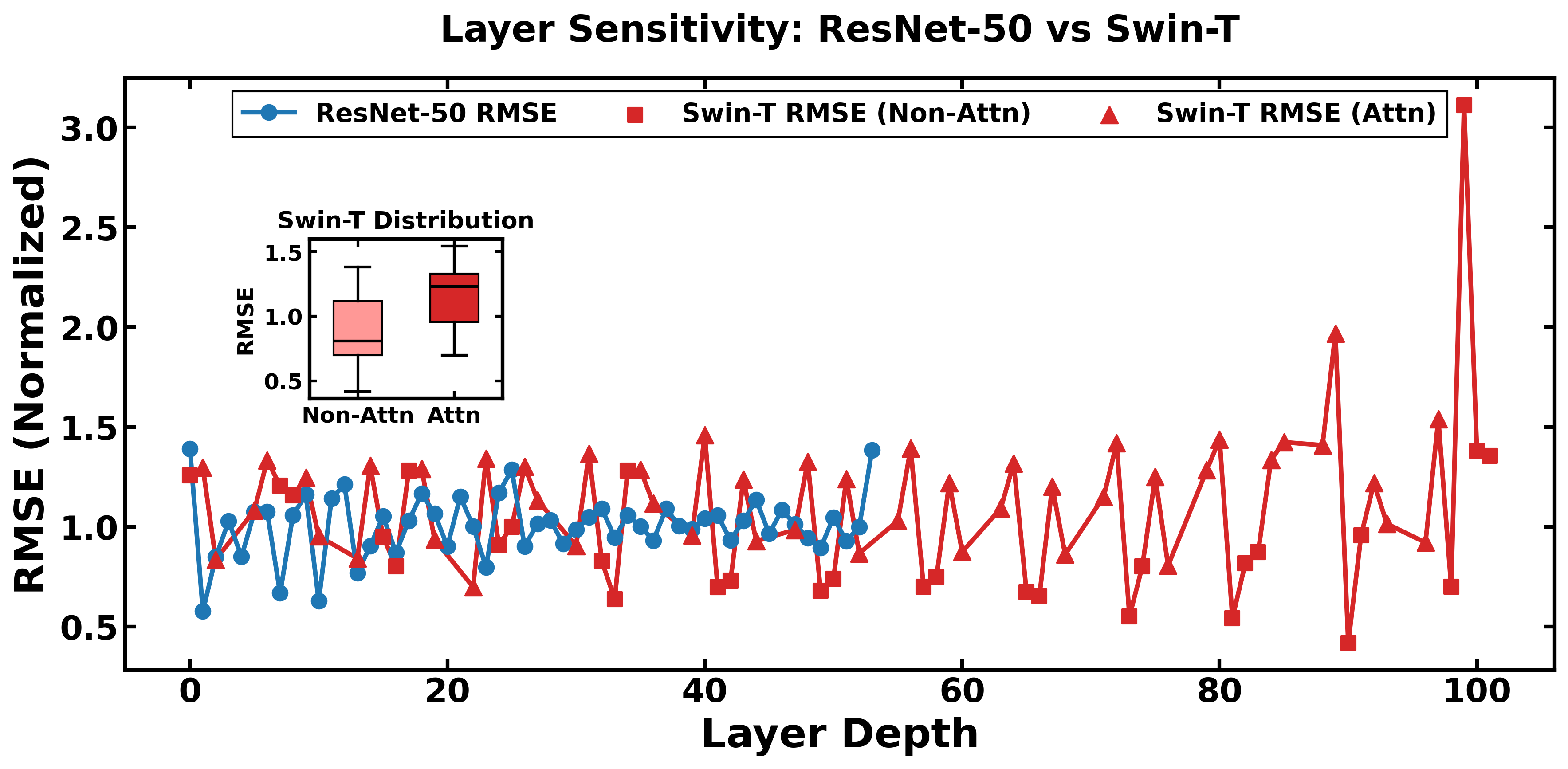}
\vspace{-4mm}
\caption{Layer-wise RMSE for ResNet-50 vs Swin-T under 4\%/2\% (HRS/LRS) D2D variation. Swin-T shows higher error variance with attention layers (triangles) exhibiting greater sensitivity than non-attention layers (squares).}
\label{fig:rmse_comparison}
\end{figure}
Fig. \ref{fig:rmse_comparison} shows the per-layer RMSE across network depth. ResNet-50 maintains consistent error across layers (mean RMSE = 1.01, $\sigma$ = 0.16) throughout its 54 layers. In contrast, Swin-T exhibits 2.25$\times$ higher error variance (mean RMSE = 1.09, $\sigma$ = 0.36) across 101 layers, indicating that certain layers are disproportionately affected by D2D variations. Notably, attention layers (marked with triangles) show 1.25$\times$ higher error than non-attention layers (marked with squares). The final layers in both models contribute disproportionately to accuracy degradation, with Swin-T's reaching an RMSE of 3.16. These findings indicate that both attention mechanism sensitivity and specific high-error outlier layers contribute to Swin-T's reduced noise tolerance.

\begin{figure}[ht]
\centering
\includegraphics[width=\linewidth]{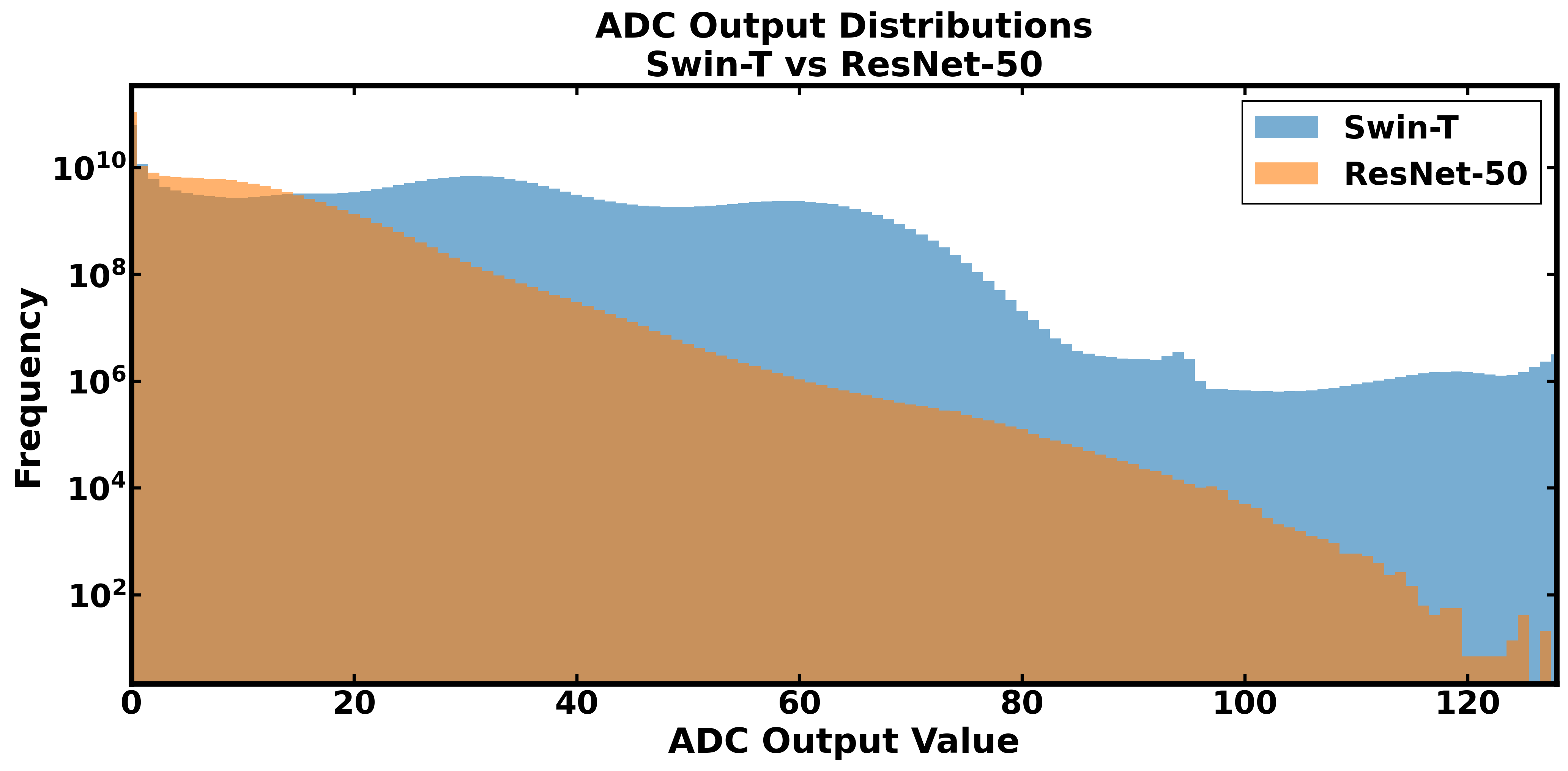}
\vspace{-4mm}
\caption{CIM ADC output distribution under ResNet-50 and Swin-T using 8b input and 8b weight with 128x128 arrays and 4\%/2\% (HRS/LRS) D2D variations. Collected from 100 images in ImageNet dataset.}
\label{fig:cim_adc_output_comparison}
\end{figure}

\subsubsection{\textbf{ADC Error Analysis}}
\begin{figure}[ht]
\centering
\includegraphics[width=\linewidth]{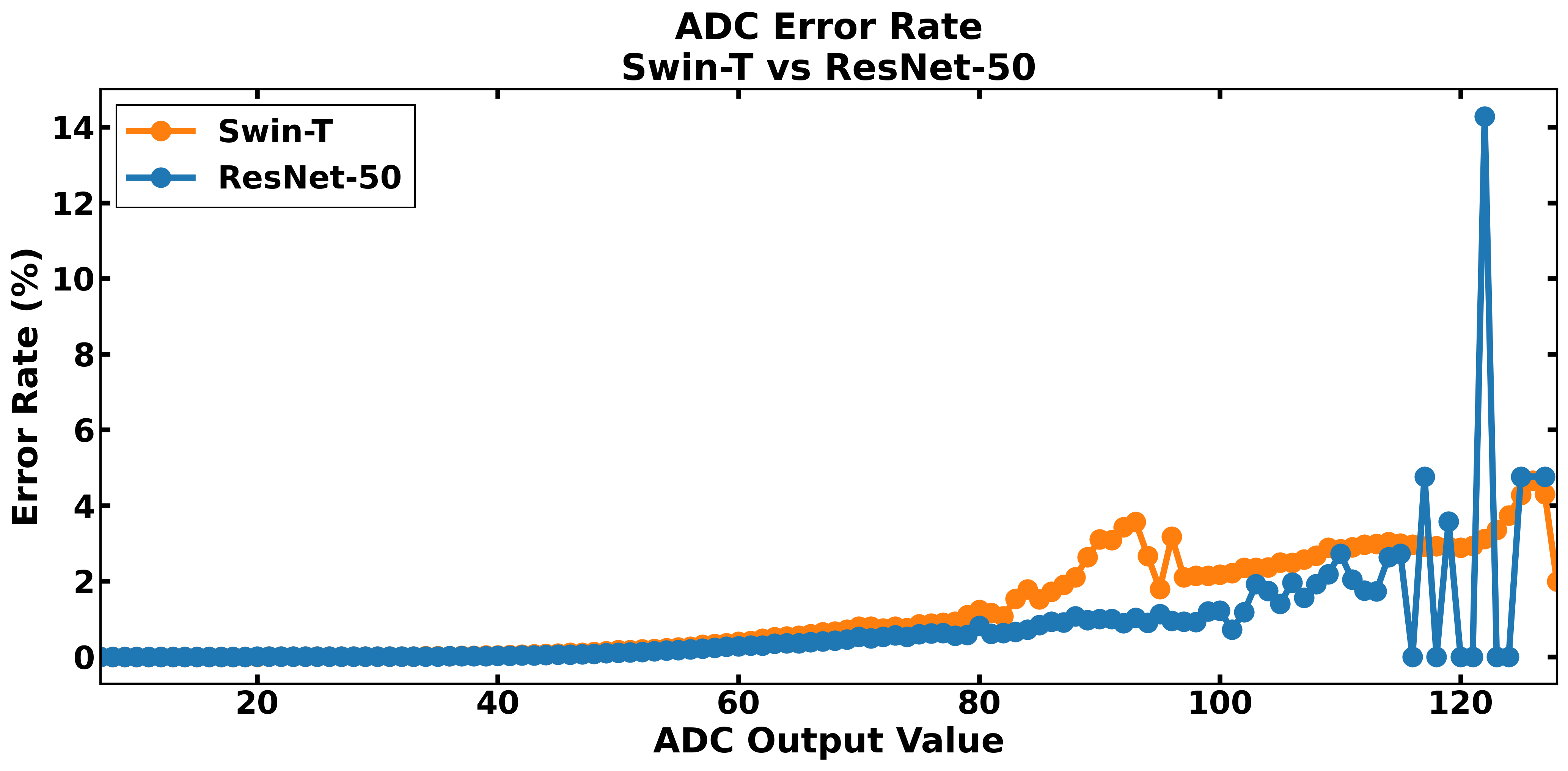}
\vspace{-4mm}
\caption{Error rate as a function of expected ADC output value for ResNet-50 and Swin-T with 128x128 arrays under 4\%/2\% (HRS/LRS) D2D variations. Collected from 100 images on the ImageNet dataset.}
\label{fig:d2d_adc_output_distribution}
\end{figure}

The layer-level RMSE analysis reveals error magnitudes after accumulating partial sums and dequantizing to floating-point. To understand the root cause of these errors, we examine errors at the ADC output stage, where integer MAC partial sums are quantized. We define error rate as the percentage of partial sums where the noisy ADC output deviates from the ideal value, computed per expected output bin. Analysis of ADC output distributions reveals that Swin-T generates significantly higher average ADC outputs compared to ResNet-50 (Fig. \ref{fig:cim_adc_output_comparison}). Larger ADC outputs experience increased error rates due to D2D variations, as shown in Fig. \ref{fig:d2d_adc_output_distribution}. Consequently, Swin-T accumulates a higher fraction of erroneous ADC outputs across all layers, contributing significantly to accuracy degradation. 

The increased error rate at higher ADC output values stems from the accumulation of device variations across activated cells. Each memory cell contributes independent D2D variation to the column current/charge. When more cells are activated (corresponding to larger partial sums), the total variance increases with the number of contributing devices. Swin-T's activation and weight distributions produce systematically higher partial sums than ResNet-50, placing a larger fraction of computations in the high-error region of Fig. \ref{fig:d2d_adc_output_distribution}.

Prior work has attributed transformer sensitivity in analog CIM primarily to MSB error amplification during bit-serial computation \cite{zhang_asim_2025}. While MSB errors do have a disproportionate impact on output magnitude, this mechanism alone does not explain why ViTs exhibit lower noise tolerance than CNNs---both architectures use identical bit-serial processing. Our analysis indicates that the distribution of ADC outputs, determined by activation sparsity and weight magnitude statistics, is the distinguishing factor. CNNs with ReLU activations produce sparser outputs concentrated at lower ADC values, while transformers with GELU activations and larger weight magnitudes shift the distribution toward higher values where error rates increase.

\subsubsection{\textbf{Mitigation Strategies}}
To improve Swin-T accuracy under analog noise, we decreased the size of the memory arrays to $32 \times 128$. This effectively reduces the ADC outputs to values less than 32, where the rate of errors is below 2\% (Fig. \ref{fig:d2d_adc_output_distribution}). As shown in Fig. \ref{d2dvariation}, this modification achieves accuracy close to the software baseline at 4\%/2\% D2D variation; however, accuracy quickly falls off at higher variation. To achieve accuracy comparable to ResNet-50 under D2D variations, we need to reduce the number of active rows in parallel to 8. The reduction in active rows does not require a new hardware configuration, as each group of 8 rows in an array are activated sequentially. We quantify the tradeoff in performance in (Table \ref{tab:parallel_read}). We find that Swin-T using $32 \times 128$ configuration achieves near parity with ResNet-50 in throughput and energy efficiency but with $5.4\times$ worse area efficiency. While comparable accuracy is achievable, CNNs remain preferable for CIM-based vision accelerators.

Alternative approaches to improve transformer resilience to ACIM noise include: (1) hardware-aware training with noise injection \cite{afm}; (2) input and weight rescaling methods to shift their distributions to fit within quantization constraints \cite{nora}; (3) Training low-rank adapters to correct errors arising from CIM noise \cite{halora}. ResNet-50's inherent advantages—weight distributions closer to zero and ReLU-induced sparsity—naturally produce lower ADC outputs and improved noise tolerance. Applying techniques to encourage these input and weight distributions in transformer architectures can bridge this robustness gap without requiring hardware modifications.

\begin{table}[htbp]
\centering
\begin{threeparttable}
\caption{22 nm RRAM ResNet-50 and Swin-T PPA under different row parallelism}
\label{tab:parallel_read}{
\renewcommand{\arraystretch}{1.4}
\begin{tabularx}{9cm}{
>{\centering\arraybackslash}X
>{\centering\arraybackslash}X
>{\centering\arraybackslash}X
>{\centering\arraybackslash}X
>{\centering\arraybackslash}X
>{\centering\arraybackslash}X
>{\centering\arraybackslash}X}
\toprule
\textbf{Model} & \textbf{Array} & \textbf{Read Rows} & \textbf{TOPS} & \textbf{TOPS / mm$^2$}  & \textbf{TOPS/W} & \textbf{FPS} \\ 
\midrule
ResNet50 & 128x128 & 128 & 4.84 & 0.0147 & 5.04 & 378      \\ 
Swin-T & 32x128 & 32 & 4.83 & 0.00272 & 4.84 & 388      \\ 
Swin-T & 32x128 & 8 & 3.46 & 0.00192 & 4.34 & 278      \\ 
\bottomrule
\end{tabularx}
}
\end{threeparttable}
\end{table}

\subsection{PPA Breakdown of Swin-T}\label{ppaBreakdownSection}
\begin{figure} [htbp]
\centering
\includegraphics[scale=0.31]{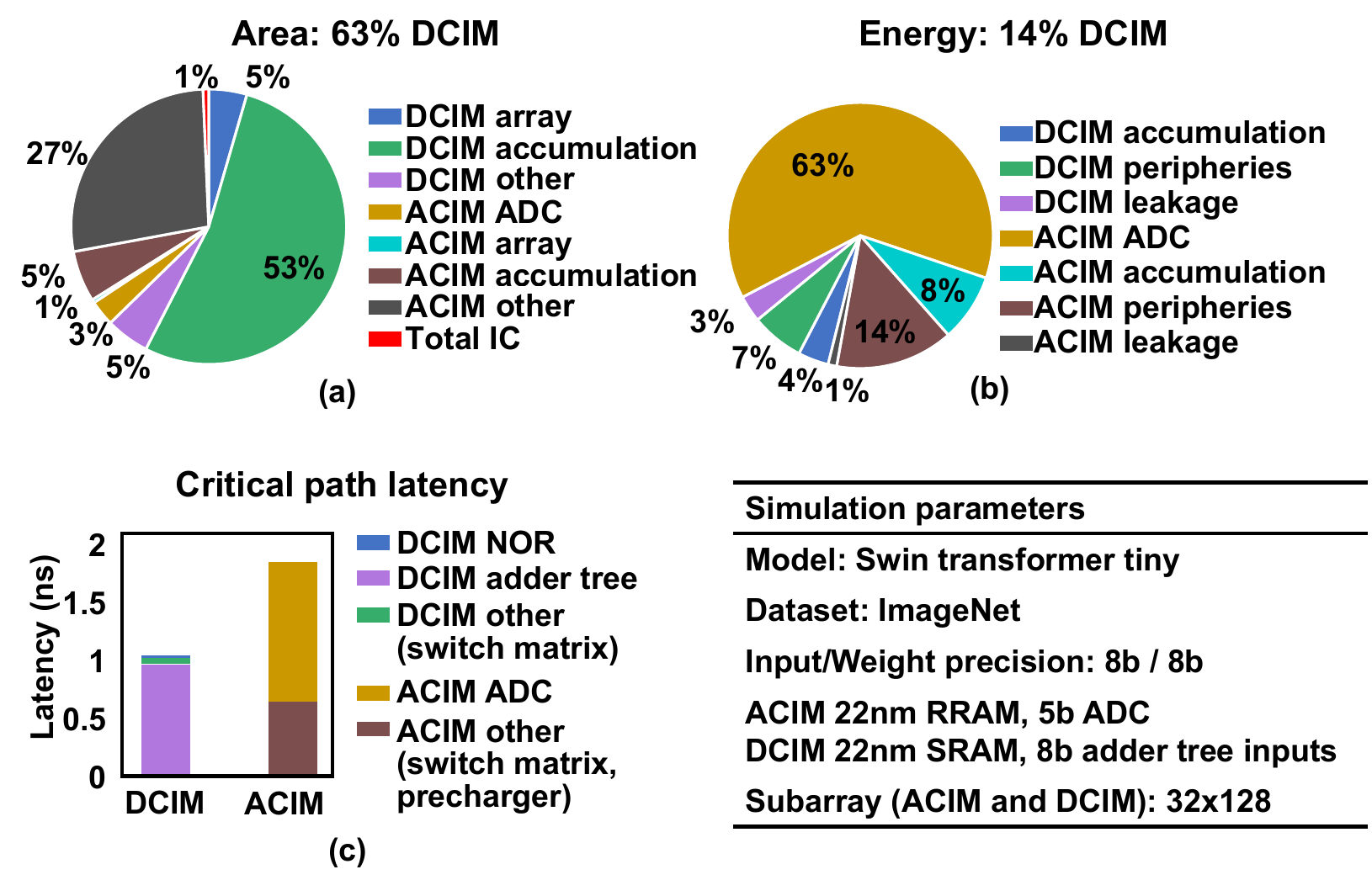}
\vspace{-2mm}
\caption{PPA breakdown of Swin-T (25M parameters). The majority of area is consumed by DCIM adder trees within memory arrays. Array size was reduced to 32x128 to decrease adder tree size. The majority of energy consumption and latency arises from ACIM ADC.}
\label{PPA}
\end{figure}

Next, we evaluate the PPA of Swin Transformer \cite{liu2021swin} on ImageNet. The linear and convolution layers are implemented in 22 nm RRAM-based ACIM while the Q,K,V matrix multiplications in attention are computed in 22 nm SRAM-based DCIM. Both CIM arrays are configured with a size of $32\times128$. Fig. \ref{PPA} presents the detailed PPA breakdown. For area, DCIM accumulation including adder trees and shift adders dominates, resulting in $1.5\times$ larger area than ACIM. However, the energy profile differs: the ADC in ACIM contributes the largest portion of energy consumption compared to significantly lower consumption of DCIM computation. This is because the linear and convolutional layers have higher arithmetic intensity than attention, requiring a higher usage of the ACIM arrays and thus higher energy consumption. This asymmetry suggests that for area-constrained designs, minimizing DCIM-computed operations (e.g., offloading some attention to external processors) may be more effective, while for energy-constrained designs, ADC optimization or reduced-precision sensing offers larger gains if accuracy can be maintained via quantization-aware tuning methods. For the critical path latency, the adder tree in DCIM and the ADC in ACIM show comparable delays. The peripheral circuits in ACIM array introduce additional overhead with longer overall latency compared to DCIM. A potential optimization for DCIM-based attention is leveraging finer-grained pipelining at the window level or head level with carefully designed dataflow and higher memory array reuse, which can reduce area while maintaining high-throughput.

\subsection{Runtime Evaluation and Simulator Comparison}
Finally, we compare NeuroSim V1.5 against existing open-source ACIM simulators (Table \ref{DNNCIMSimulators}).
CrossSim \cite{xiao_tianyao_crosssim_2021} excels at device-level physical simulation with high fidelity but higher runtime, especially for complex noise. AIHWKit \cite{gallo_using_2023} focuses on noise-aware training with phase-change memory. NeuroSim V1.4 \cite{lee_neurosim_2024} introduced integrated PPA and some noise modeling but had slower inference and narrower network support. NeuroSim V1.5 distinguishes itself by integrating flexible, gpu-accelerated noise modeling, and vision transformer support, with circuit-level PPA estimation. This enables a wider range of noise, quantization, and circuit parameters to be explored than in other open-source simulators.

\begin{table}[ht]
\centering
\begin{threeparttable}
\caption{Comparison of DNN ACIM Simulators}
\label{DNNCIMSimulators}
\renewcommand{\arraystretch}{1.4}
\begin{tabularx}{9cm}{
>{\centering\arraybackslash}m{1.8cm}  
>{\centering\arraybackslash}m{0.4cm} 
>{\centering\arraybackslash}m{0.4cm} 
>{\centering\arraybackslash}m{0.4cm} 
>{\centering\arraybackslash}m{0.4cm} 
>{\centering\arraybackslash}m{0.3cm} 
>{\centering\arraybackslash}m{0.3cm}
>{\centering\arraybackslash}m{0.4cm} 
>{\centering\arraybackslash}m{0.4cm}
}
\toprule
\textbf{DNN CIM Simulator} 
& \multicolumn{2}{c}{\textbf{\begin{tabular}[c]{@{}c@{}} CrossSim \\ \cite{xiao_tianyao_crosssim_2021} \end{tabular}}} 
& \multicolumn{2}{c}{\textbf{\begin{tabular}[c]{@{}c@{}} AIHWKit \\ \cite{gallo_using_2023} \end{tabular}}} 
& \multicolumn{2}{c}{\textbf{\begin{tabular}[c]{@{}c@{}} NeuroSim \\ V1.4 \cite{lee_neurosim_2024}\end{tabular}}} 
& \multicolumn{2}{c}{\textbf{\begin{tabular}[c]{@{}c@{}} NeuroSim \\ V1.5\end{tabular}}}  \\
\midrule
Supported Devices
& \multicolumn{2}{c}{\renewcommand{\arraystretch}{1.1}%
\begin{tabular}[c]{@{}c@{}}PCM,\\ RRAM, \\ EcRAM,\\ DRAM\end{tabular}} 
& \multicolumn{2}{c}{\renewcommand{\arraystretch}{1.1}%
\begin{tabular}[c]{@{}c@{}}RRAM,\\ PCM,\\ Flash\end{tabular}} 
& \multicolumn{2}{c}{\renewcommand{\arraystretch}{1.1}%
\begin{tabular}[c]{@{}c@{}}SRAM,\\ RRAM,\\ FeFET\end{tabular}} 
& \multicolumn{2}{c}{\renewcommand{\arraystretch}{1.1}%
\begin{tabular}[c]{@{}c@{}}SRAM,\\ RRAM,\\ FeFET,\\ nvCap\end{tabular}} \\
Supported Network Types
& \multicolumn{2}{c}{\renewcommand{\arraystretch}{1.1}%
\begin{tabular}[c]{@{}c@{}}Linear,\\ Conv \end{tabular}} 
& \multicolumn{2}{c}{\renewcommand{\arraystretch}{1.1}%
\begin{tabular}[c]{@{}c@{}}Linear,\\ Conv, \\ Recurrent,\\  Transformer\end{tabular}} 
& \multicolumn{2}{c}{\renewcommand{\arraystretch}{1.1}%
\begin{tabular}[c]{@{}c@{}}Linear,\\ Conv, \\ Recurrent \end{tabular}} 
& \multicolumn{2}{c}{\renewcommand{\arraystretch}{1.1}%
\begin{tabular}[c]{@{}c@{}}Linear,\\ Conv, \\ Recurrent,\\  Transformer\end{tabular}} \\
Bit Slicing Support & \multicolumn{2}{c}{Yes} & \multicolumn{2}{c}{No} & \multicolumn{2}{c}{Yes} & \multicolumn{2}{c}{Yes}  \\
PPA Support & \multicolumn{2}{c}{No} & \multicolumn{2}{c}{No} & \multicolumn{2}{c}{Yes} & \multicolumn{2}{c}{Yes}  \\
 \bottomrule
\end{tabularx}

\end{threeparttable}
\end{table}

Table \ref{RuntimeComparison1p41p5} details runtime scaling with precision and noise configuration. NeuroSim V1.5 achieves 0.95 ms/image baseline for VGG8/CIFAR-10, increasing to 4.1 ms/image with 1b cells due to bit-slicing overhead. Device noise adds negligible cost; per-MAC output noise increases runtime to 5.1 ms/image.

\begin{figure} [htbp]
\centering
\includegraphics[scale=0.4]{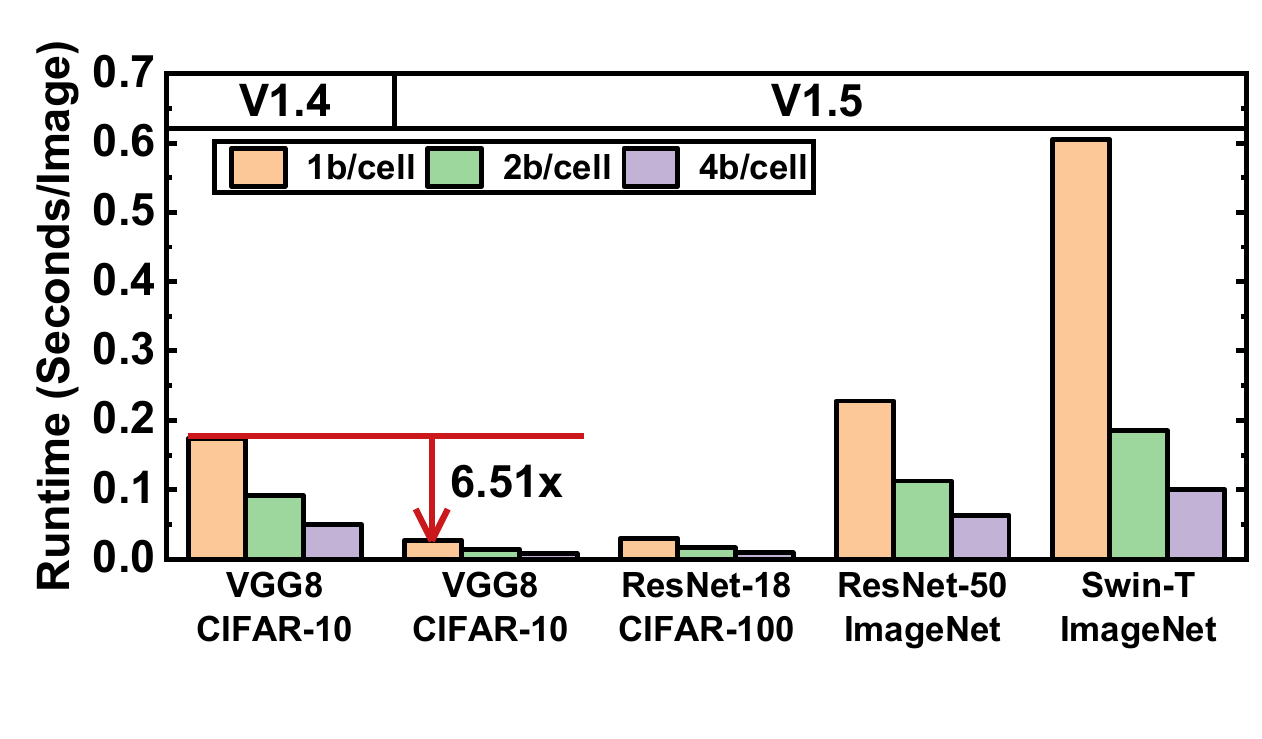}
\vspace{-2mm}
\caption{Inference runtime of different neural network algorithms using V1.4 and V1.5 across 1b/2b/4b RRAM cell states without device non-idealities and noise.}
\label{runTime}
\end{figure}

Runtime improvements stem from redesigned GPU-accelerated tensor operations that simulate all arrays and sample noise distributions in parallel. As Fig. \ref{runTime} shows, these optimizations deliver up to 6.51$\times$ speedup over V1.4.

\begin{table}[htbp]
\centering
\begin{threeparttable}
\caption{Inference Runtime (Seconds/Image) Comparison of VGG8 on CIFAR-10 (8b activation, 8b weight) between NeuroSim V1.4 and NeuroSim V1.5}
\label{RuntimeComparison1p41p5}
\renewcommand{\arraystretch}{1.4}
\begin{tabular}{|>{\centering\arraybackslash}m{3cm}|>{\centering\arraybackslash}m{2.2cm}|>{\centering\arraybackslash}m{2.2cm}|}
\hline
\renewcommand{\arraystretch}{1.1}%
\textbf{\begin{tabular}[c]{@{}c@{}}DAC Precision/ \\ MLC/Noise \end{tabular}}&
\renewcommand{\arraystretch}{1.1}%
\textbf{\begin{tabular}[c]{@{}c@{}}NeuroSim \\ V1.4 \cite{lee_neurosim_2024} \end{tabular}} &
\renewcommand{\arraystretch}{1.1}%
\textbf{\begin{tabular}[c]{@{}c@{}}NeuroSim \\ V1.5 \end{tabular}}\\
\hline
8b/8b/None & -- & 0.00095 \\ \hline
8b/1b/None & -- & 0.0041 \\ \hline
1b/8b/None & 0.042 & 0.0048 \\ \hline
1b/1b/None & 0.17 & 0.027 \\ \hline
8b/8b/Device Noise & -- & 0.00095 \\ \hline
8b/1b/Device Noise & -- & 0.0041 \\ \hline
1b/8b/Device Noise & 0.042 & 0.0048 \\ \hline
1b/1b/Device Noise & 0.17 & 0.027 \\ \hline
8b/8b/Output noise & -- & 0.0011 \\ \hline
8b/1b/Output noise & -- & 0.0051 \\ \hline
1b/8b/Output noise & -- & 0.0059 \\ \hline
1b/1b/Output noise & -- & 0.039,0.085* \\ \hline
\end{tabular}
\begin{tablenotes}
\item $128\times128$ array size
\item * 0.039: same noise on each MAC output, 0.085: individual noise on each MAC output
\end{tablenotes}
\end{threeparttable}
\end{table}

Table \ref{RuntimeComparisonCrossAIHW1p5} demonstrates V1.5's efficient noise modeling on larger networks. Compared to CrossSim, which shows significant slowdown when incorporating device noise, V1.5's statistical approach adds minimal overhead - just 1.3$\times$ for uniform noise and 3.1$\times$ for output-dependent noise.

These advances improve NeuroSim's functionality as a design space exploration platform across memory devices, computing methods, and technology nodes. Evaluating both network accuracy and performance is essential for optimal CIM hardware design, as ignoring either metric provides an incomplete picture and can exaggerate throughput and energy-efficiency. 

\begin{table}[htbp]
\centering
\begin{threeparttable}
\caption{Inference Runtime (Seconds/Image) Comparison of ResNet-50 on ImageNet (8b activation, 8b weight) between CrossSim, AIHWKit, and NeuroSim V1.5}
\label{RuntimeComparisonCrossAIHW1p5}
\renewcommand{\arraystretch}{1.4}
\begin{tabular}{|>{\centering\arraybackslash}m{2.6cm}|>{\centering\arraybackslash}m{1.4cm}|>{\centering\arraybackslash}m{1.4cm}|>{\centering\arraybackslash}m{1.5cm}|}
\hline
\renewcommand{\arraystretch}{1.1}%
\textbf{\begin{tabular}[c]{@{}c@{}}DAC Precision/ \\ MLC/Noise \end{tabular}}&
\renewcommand{\arraystretch}{1.1}%
\textbf{\begin{tabular}[c]{@{}c@{}}CrossSim\\ \cite{xiao_tianyao_crosssim_2021} \end{tabular}} &
\renewcommand{\arraystretch}{1.1}%
\textbf{\begin{tabular}[c]{@{}c@{}}AIHWKit\\ \cite{gallo_using_2023} \end{tabular}} &
\renewcommand{\arraystretch}{1.1}%
\textbf{\begin{tabular}[c]{@{}c@{}}NeuroSim \\ V1.5 \end{tabular}}\\
\hline
8b/8b/None & 0.3 & 0.0025 & 0.0085 \\ \hline
8b/1b/None & 1.1 & -- & 0.033 \\ \hline
1b/8b/None & 1 & -- & 0.04 \\ \hline
1b/1b/None & 5.8 & -- & 0.24 \\ \hline
8b/8b/Device Noise & 9.8 & 0.003 & 0.0085 \\ \hline
8b/1b/Device Noise & 200 & -- & 0.033 \\ \hline
1b/8b/Device Noise & 45 & -- & 0.041 \\ \hline
1b/1b/Device Noise & 1220 & -- & 0.24 \\ \hline
8b/8b/Output noise & -- & 0.0025 & 0.0093 \\ \hline
8b/1b/Output noise & -- & -- & 0.039 \\ \hline
1b/8b/Output noise & -- & -- & 0.047 \\ \hline
1b/1b/Output noise & -- & -- & 0.29,0.61* \\ \hline
\end{tabular}
\begin{tablenotes}
\item $128\times128$ array size
\item * 0.29: same noise on each MAC output, 0.61: individual noise on each MAC output
\end{tablenotes}
\end{threeparttable}
\end{table}

\section{Conclusion and Future Work}\label{conclusion}
The growing complexity of AI models places increasing demands on CIM accelerator design, requiring tools that can efficiently evaluate both accuracy and hardware performance. This work presents NeuroSim V1.5, which combines accuracy analysis with power, performance, and area estimation. A key contribution is the heterogeneous ACIM/DCIM architecture extending CIM support to transformer models---increasingly important as transformers dominate AI workloads from vision to language.

The framework's integration with TensorRT enables pre-trained network evaluation without modification. Our flexible noise modeling supports both device-level analysis and efficient MAC-level noise injection, validated against published SPICE/silicon measurements. These capabilities, with up to 6.5$\times$ faster runtime than NeuroSim V1.4, enable rapid design space exploration.

Through our case studies with various device technologies and neural network topologies, we demonstrate how NeuroSim V1.5 can provide insights into CIM design trade-offs:
\begin{enumerate}
    \item Pareto-optimal designs converge around the 5-7 bit ADC configuration, underscoring the importance of ADC overhead in ACIM architectures.
    \item Heterogeneous ACIM/DCIM architectures enable transformer acceleration; however, limiting DCIM circuits to ACIM-compatible technology nodes reduces area efficiency. Heterogeneous chiplet integration with separate ACIM and DCIM dies could offer a solution.
    \item ViTs exhibit lower noise tolerance and area efficiency than CNNs of comparable size. For vision applications where both architectures are viable, CNNs remain better suited for CIM hardware.
\end{enumerate}

Looking forward, several enhancements are planned: lookup table-based noise modeling for complex effects like IR drop, generalizing the PPA framework for flexible system-level exploration similar to CIMLoop, and incorporating time-domain input encoding for broader CIM coverage.

Extending NeuroSim to support large language models is a primary future direction, as transformers are the de-facto architecture for these workloads. LLMs present unique challenges requiring GB-level parameter storage---addressable through multi-die CIM integration or process-in-memory (PIM) architectures with CMOS compute near high-bandwidth memory. We have initiated PIM efforts through our work on 3D-stacked DRAM architectures \cite{sharda_accelerator_2024, hsu_dram_2025}.

\newpage

\begin{IEEEbiography}
[{\includegraphics[width=1in,height=1.25in,clip,keepaspectratio]{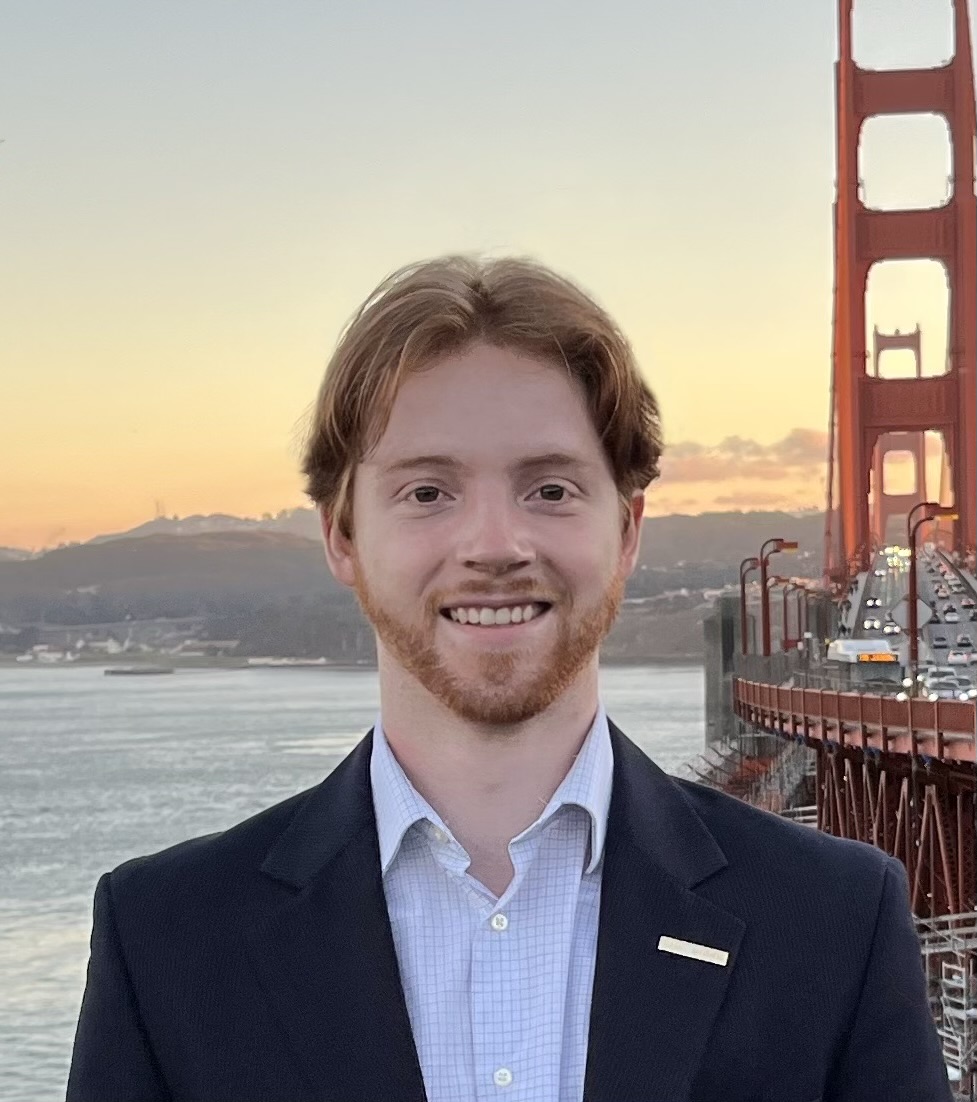}}]{James Read} (Graduate Student Member, IEEE) received his B.S. degree in computer engineering from the University of Michigan, Ann Arbor, MI, in 2021. He is currently a Ph.D. candidate in electrical and computer engineering with the Georgia Institute of Technology, Atlanta, GA, USA. His research interests include in-memory computing with emerging non-volatile memory devices, architecture and security of deep learning accelerators, and neuromorphic computing.
\end{IEEEbiography}

\vspace{-30pt}

\begin{IEEEbiography}
[{\includegraphics[width=1in,height=1.25in,clip,keepaspectratio]{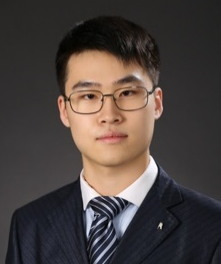}}]{Ming-Yen Lee} (Graduate Student Member, IEEE) received the B.S. and M.S. degree in the Department of Electronic Engineering from Tsinghua University, Beijing, China, in 2021 and 2024, respectively. He is currently a PhD student in the School of Electrical and Computer Engineering, Georgia Institute of Technology, Atlanta, GA, USA. His research interests mainly include compute-in-memory and software-hardware co-design for emerging AI model.
\end{IEEEbiography}

\vspace{-30pt}

\begin{IEEEbiography}
[{\includegraphics[width=1in,height=1.25in,clip,keepaspectratio]{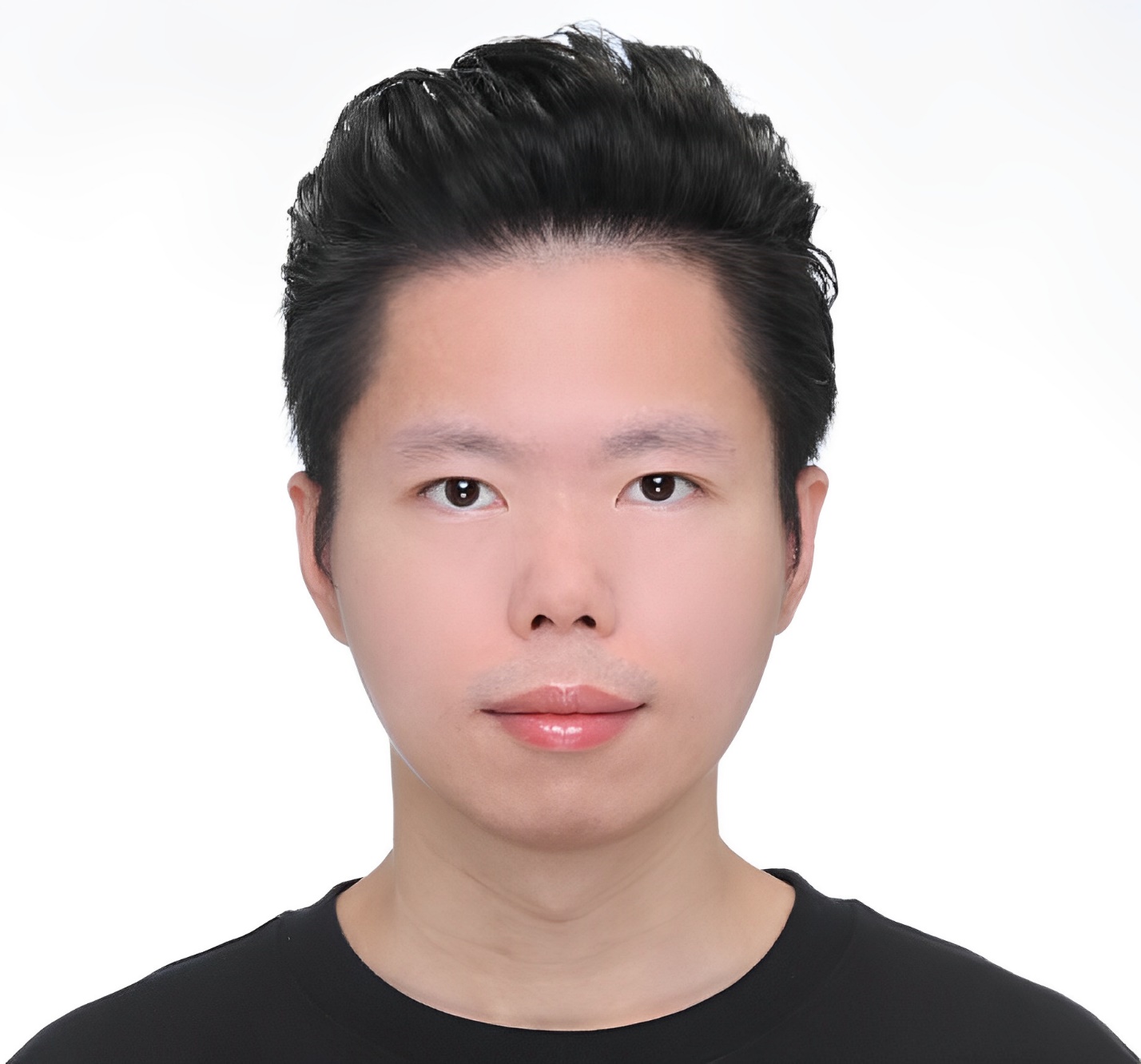}}]{Wei-Hsing Huang} Wei-Hsing Huang received the B.S. degree in electrical engineering from the National Chung Cheng University, Chiayi, Taiwan, in 2017, and the M.S. degree in electrical engineering and computer science from the National Tsing Hua University, Hsinchu, Taiwan, in 2019. He is currently a Research Assistant in electrical and computer engineering with Georgia Institute of Technology, Atlanta, GA, USA. His current research interests include deep learning algorithms and hardware-software co-design for deep learning.
\end{IEEEbiography}

\vspace{-30pt}

\begin{IEEEbiography}
[{\includegraphics[width=1in,height=1.25in,clip,keepaspectratio]{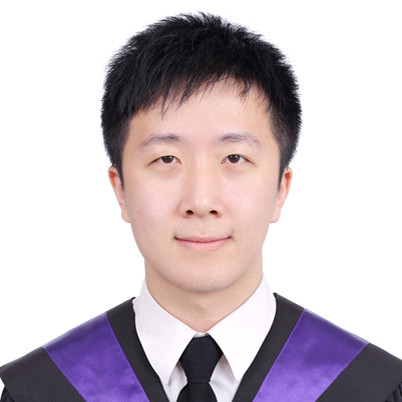}}]{Yuan-Chun Luo} (Graduate Student Member, IEEE) received the B.S. degree in electrical engineering from the National Tsing Hua University, Taiwan, in 2018. He is currently pursuing the Ph.D. degree in electrical and computer engineering with the Georgia Institute of Technology, Atlanta, GA, USA. His current research interests include device-circuit co- design for emerging non-volatile memories.
\end{IEEEbiography}

\vspace{-30pt}

\begin{IEEEbiography}
[{\includegraphics[width=1in,height=1.25in,clip,keepaspectratio]{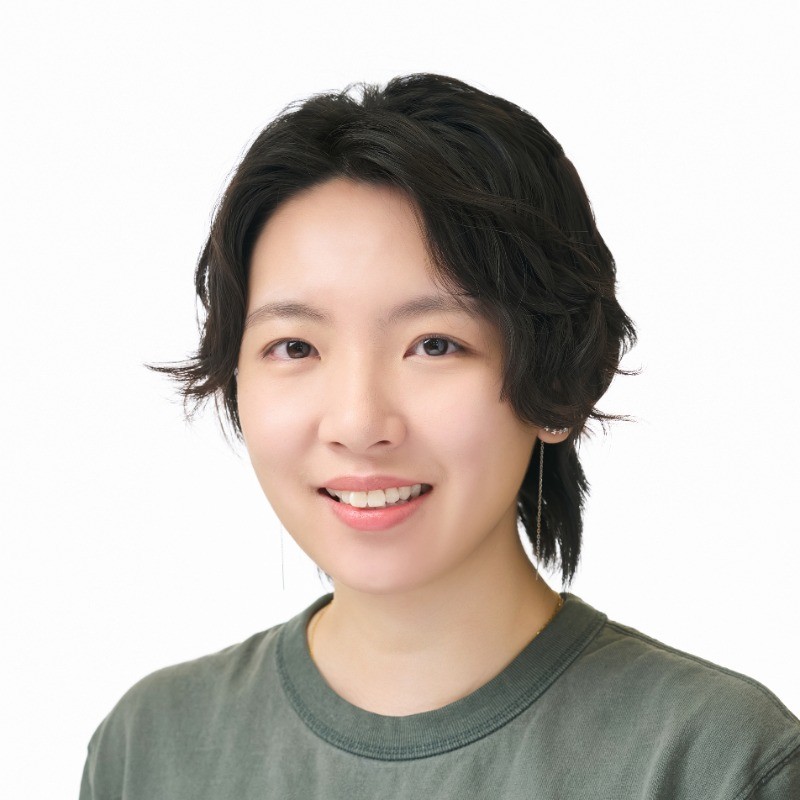}}]{Anni Lu} (Graduate Student Member, IEEE) received the B.S. degree in electronic information engineering from Tianjin University, China, in 2019. She is currently pursuing the Ph.D. degree in electrical and computer engineering with the Georgia Institute of Technology, Atlanta, GA, USA. Her current research interests include algorithm-hardware co- design and device-to-system benchmarking framework for hardware accelerators of deep learning and beyond.
\end{IEEEbiography}

\vspace{-30pt}

\begin{IEEEbiography}
[{\includegraphics[width=1in,height=1.25in,clip,keepaspectratio]{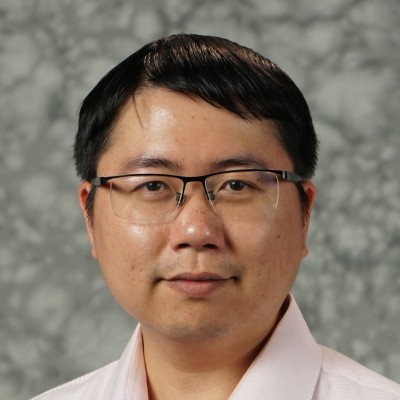}}]{Shimeng Yu} (Fellow, IEEE) is Dean's Professor of electrical and computer engineering at Georgia Institute of Technology. He received his B.S. from Peking University in 2009 and Ph.D. from Stanford University in 2013. He is an IEEE Fellow for contributions in non-volatile memories and CIM. His research focuses on memory technologies and 3D integration for AI hardware. His honors include the NSF CAREER Award, IEEE EDS Early Career Award, ACM SIGDA Outstanding New Faculty Award, and SRC Technical Excellence Award.
\end{IEEEbiography}

\vfill


\begin{thebibliography}{1}
\bibliographystyle{IEEEtran}
\bibitem{tripp_measuring_2024} C. E. Tripp, et al., ``Measuring the energy consumption and efficiency of deep neural networks: An empirical analysis and design recommendations,'' arXiv:2403.08151, Mar. 2024.
\bibitem{williams_roofline_2009} S. Williams, et al., ``Roofline: An insightful visual performance model for multicore architectures,'' \textit{Commun. ACM}, vol. 52, no. 4, pp. 65--76, Apr. 2009.
\bibitem{chih_164_2021} Y.-D. Chih, et al., ``16.4 An 89 TOPS/W and 16.3TOPS/mm$^2$ all-digital SRAM-based full-precision compute-in memory macro in 22nm for machine-learning edge applications,'' in \textit{IEEE Int. Solid-State Circuits Conf. (ISSCC)}, Feb. 2021, vol. 64, pp. 252--254.
\bibitem{yu_neuro-inspired_2018} S. Yu, ``Neuro-inspired computing with emerging non-volatile memories,'' \textit{Proc. IEEE}, vol. 106, no. 2, pp. 260--285, Feb. 2018.
\bibitem{verma_-memory_2019} N. Verma, et al., ``In-memory computing: Advances and prospects,'' \textit{IEEE Solid-State Circuits Mag.}, vol. 11, no. 3, pp. 43--55, Summer 2019.
\bibitem{jain_132_2019} P. Jain, et al., ``13.2 A 3.6Mb 10.1Mb/mm$^2$ Embedded non-volatile ReRAM macro in 22nm FinFET technology with adaptive forming/set/reset schemes yielding down to 0.5V with sensing time of 5ns at 0.7V,'' in \textit{IEEE Int. Solid-State Circuits Conf. (ISSCC)}, Feb. 2019, pp. 212--214.
\bibitem{9755965} W. Li, et al., ``A 40nm MLC-RRAM compute-in-memory macro with sparsity control, on-chip write-verify, and temperature-independent ADC references,'' \textit{IEEE J. Solid-State Circuits}, vol. 57, no. 9, pp. 2868--2877, Sep. 2022.
\bibitem{cheng-xin_xue_241_2019} C.-X. Xue, et al., ``24.1 a 1MB multibit ReRAM computing-in-memory macro with 14.6ns parallel MAC computing time for CNN based AI edge processors,'' in \textit{IEEE Int. Solid-State Circuits Conf. (ISSCC)}, Feb. 2019, pp. 388--390.
\bibitem{qi_liu_332_2020} Q. Liu, et al., ``33.2 a fully integrated analog reram based 78.4tops/w compute-in-memory chip with fully parallel mac computing,'' in \textit{IEEE Int. Solid-State Circuits Conf. (ISSCC)}, Feb. 2020, pp. 500--502.
\bibitem{le_gallo_64-core_2023} M. Le Gallo, et al., ``A 64-core mixed-signal in-memory compute chip based on phase-change memory for deep neural network inference,'' \textit{Nat. Electron.}, vol. 6, pp. 680--693, Sep. 2023.
\bibitem{vinay_joshi_accurate_2020} V. Joshi, M. Le Gallo, S. Haefeli, I. Boybat, S. R. Nandakumar, C. Piveteau, et al., ``Accurate deep neural network inference using computational phase-change memory,'' \textit{Nat. Commun.}, vol. 11, no. 1, p. 2473, May 2020.
\bibitem{mulaosmanovic_novel_2017} H. Mulaosmanovic, et al., ``Novel ferroelectric FET based synapse for neuromorphic systems,'' in \textit{Symp. VLSI Technol. Dig. Tech. Papers}, Jun. 2017, pp. T182--T183.
\bibitem{jerry_ferroelectric_2017} M. Jerry, et al., ``Ferroelectric FET analog synapse for acceleration of deep neural network training,'' in \textit{IEEE Int. Electron Devices Meeting (IEDM)}, Dec. 2017, pp. 6.2.1--6.2.4.
\bibitem{de_demonstration_2022} S. De, et al., ``Demonstration of multiply-accumulate operation with 28 nm FeFET crossbar array,'' \textit{IEEE Electron Device Lett.}, vol. 43, no. 12, pp. 2081--2084, Dec. 2022.
\bibitem{eFlash} E. Choi et al., "A 333TOPS/W Logic-Compatible Multi-Level Embedded Flash Compute-In-Memory Macro with Dual-Slope Computation," 2023 IEEE Custom Integrated Circuits Conference (CICC), San Antonio, TX, USA, 2023, pp. 1-2.
\bibitem{10272016} S. Yu, et al., ``Nonvolatile capacitive synapse: Device candidates for charge domain compute-in-memory,'' \textit{IEEE Electron Devices Mag.}, vol. 1, no. 2, pp. 23--32, Jun. 2023.
\bibitem{luo_experimental_2021} Y.-C. Luo, et al., ``Experimental demonstration of non-volatile capacitive crossbar array for in-memory computing,'' in \textit{IEEE Int. Electron Devices Meeting (IEDM)}, Dec. 2021, pp. 35.5.1--35.5.4.
\bibitem{10237236} T.-H. Kim, et al., ``Tunable non-volatile gate-to-source/drain capacitance of fefet for capacitive synapse,'' \textit{IEEE Electron Device Lett.}, vol. 44, no. 10, pp. 1628--1631, Oct. 2023.
\bibitem{sinangil_7-nm_2021} M. E. Sinangil, et al., ``A 7-nm compute-in-memory SRAM macro supporting multi-bit input, weight and output and achieving 351 TOPS/W and 372.4 GOPS,'' \textit{IEEE J. Solid-State Circuits}, vol. 56, no. 1, pp. 188--198, Jan. 2021.
\bibitem{8662392} X. Si, et al., ``24.5 a twin-8T SRAM computation-in-memory macro for multiple-bit cnn-based machine learning,'' in \textit{IEEE Int. Solid-State Circuits Conf. (ISSCC)}, Feb. 2019, pp. 396--398.
\bibitem{9062995} X. Si, et al., ``15.5 a 28nm 64kb 6T SRAM computing-in-memory macro with 8b mac operation for ai edge chips,'' in \textit{IEEE Int. Solid-State Circuits Conf. (ISSCC)}, Feb. 2020, pp. 246--248.
\bibitem{valavi_64-tile_2019} H. Valavi, et al., ``A 64-tile 2.4-Mb in-memory-computing CNN accelerator employing charge-domain compute,'' \textit{IEEE J. Solid-State Circuits}, vol. 54, no. 6, pp. 1789--1799, Jun. 2019.
\bibitem{lee_fully_2021} J. Lee, et al., ``Fully row/column-parallel in-memory computing SRAM macro employing capacitor-based mixed-signal computation with 5-b Inputs,'' in \textit{Symp. VLSI Circuits Dig. Tech. Papers}, Jun. 2021, pp. 1--2.
\bibitem{changhyuck_sung_effect_2018} C. Sung, S. Lim, H. Kim, T. Kim, K. Moon, J. Song, et al., ``Effect of conductance linearity and multi-level cell characteristics of TaOx-based synapse device on pattern recognition accuracy of neuromorphic system,'' \textit{Nanotechnology}, vol. 29, no. 11, p. 115203, Mar. 2018.
\bibitem{wei_wu_methodology_2018} W. Wu, et al., ``A methodology to improve linearity of analog rram for neuromorphic computing,'' in \textit{Symp. VLSI Technol. Dig. Tech. Papers}, Jun. 2018, pp. 103--104.
\bibitem{xiaoyu_sun_impact_2019} X. Sun, et al., ``Impact of Non-ideal characteristics of resistive synaptic devices on implementing convolutional neural networks,'' \textit{IEEE J. Emerg. Sel. Topics Circuits Syst.}, vol. 9, no. 3, pp. 570--579, Sep. 2019.
\bibitem{luo_cross-layer_2024} Y.-C. Luo, et al., ``A cross-layer framework for design space and variation analysis of non-volatile ferroelectric capacitor-based compute-in-memory accelerators,'' in \textit{Proc. Asia South Pacific Design Autom. Conf. (ASP-DAC)}, Jan. 2024, pp. 159--164.
\bibitem{matthew_spear_impact_2023} M. Spear, et al., ``The impact of analog-to-digital converter architecture and variability on analog neural network accuracy,'' \textit{IEEE J. Explor. Solid-State Comput. Devices Circuits}, vol. 9, pp. 61--70, Jan. 2023.
\bibitem{huang_hardware-aware_2023} S. Huang, et al., ``Hardware-aware quantization/mapping strategies for compute-in-memory accelerators,'' \textit{ACM Trans. Design Autom. Electron. Syst.}, vol. 28, no. 3, pp. 1--23, Mar. 2023.
\bibitem{xiao_tianyao_crosssim_2021} T. Xiao, et al., ``Crosssim: gpu-accelerated simulation of analog neural networks,'' Sandia National Lab.(SNL-NM), Albuquerque, NM (United States), Tech. Rep. SAND2021-2 CrossSim, 2021. [Online]. Available: https://github.com/sandialabs/cross-sim
\bibitem{rasch_flexible_2021} M. J. Rasch, et al., ``A flexible and fast PyTorch toolkit for simulating training and inference on analog crossbar arrays,'' in \textit{Proc. IEEE Int. Conf. Artif. Intell. Circuits Syst. (AICAS)}, Jun. 2021, pp. 1--4.
\bibitem{andrulis_cimloop_2024} T. Andrulis, et al., ``CiMLoop: a flexible, accurate, and fast compute-in-memory modeling tool,'' arXiv:2405.07259, May 2024.
\bibitem{read_enabling_2023} J. Read et al., ``Enabling long-term robustness in RRAM-based compute-in-memory edge devices,'' in \textit{Proc. IEEE Int. Symp. Circuits Syst. (ISCAS)}, May 2023, pp. 1--5.
\bibitem{yixin_multilevel_2024} Y. Xu, et al., ``Multi-level cell sensing inspired robust and energy-efficient charge domain compute-in-memory array with ferroelectric fet,'' in \textit{IEEE Int. Electron Devices Meeting (IEDM)}, Dec. 2024.
\bibitem{hur_nonvolatile_2022} J. Hur, et al., ``Nonvolatile capacitive crossbar array for in-memory computing,'' \textit{Adv. Intell. Syst.}, vol. 4, no. 8, Aug. 2022, Art. no. 2100269.
\bibitem{kim_capacitive_2024} C.-K. Kim et al., ``Capacitive synaptor with gate surrounding semiconductor pillar structure and overturned charge injection for compute-in-memory,'' \textit{Adv. Intell. Syst.}, Aug. 2024, Art. no. 2400371.
\bibitem{9896828} J.-W. Su, et al., ``A 8-b-precision 6T SRAM computing-in-memory macro using segmented-bitline charge-sharing scheme for ai edge chips,'' \textit{IEEE J. Solid-State Circuits}, vol. 58, no. 3, pp. 877--892, Mar. 2023.
\bibitem{jia_151_2021} H. Jia, et al., ``15.1 A programmable neural-network inference accelerator based on scalable in-memory computing,'' in \textit{IEEE Int. Solid-State Circuits Conf. (ISSCC)}, Feb. 2021, vol. 64, pp. 236--238.
\bibitem{peng_dnnneurosim_2019} X. Peng, et al., ``DNN+NeuroSim: an end-to-end benchmarking framework for compute-in-memory accelerators with versatile device technologies,'' in \textit{IEEE Int. Electron Devices Meeting (IEDM)}, Dec. 2019, pp. 11.3.1--11.3.4.
\bibitem{wan_compute--memory_2022} W. Wan, R. Kubendran, C. Schaefer, S. B. Eryilmaz, W. Zhang, D. Wu, et al., ``A compute-in-memory chip based on resistive random-access memory,'' \textit{Nature}, vol. 608, no. 7923, pp. 504--512, Aug. 2022.
\bibitem{aabrar_thousand_2022} K. A. Aabrar, et al., ``A thousand state superlattice(SL) FEFET analog weight cell,'' in \textit{Symp. VLSI Technol. Circuits Dig. Tech. Papers}, Jun. 2022, pp. 242--243.
\bibitem{pwm_nand} S.-T. Lee, et al., ``Neuromorphic computing using NAND flash memory architecture with pulse width modulation scheme,'' \textit{Front. Neurosci.}, vol. 14, Aug. 2020, Art. no. 773.
\bibitem{timaq} J. Yang et al., ``TIMAQ: A time-domain computing-in-memory-based processor using predictable decomposed convolution for arbitrary quantized DNNs,'' \textit{IEEE J. Solid-State Circuits}, Oct. 2021.
\bibitem{yuyao} Y. Kong, et al., ``Evaluation platform of time-domain computing-in-memory circuits,'' \textit{IEEE Trans. Circuits Syst. II, Exp. Briefs}, vol. 70, no. 3, pp. 1174--1178, Mar. 2023.
\bibitem{jiang_enna_2023} H. Jiang, et al., ``ENNA: an efficient neural network accelerator design based on ADC-free compute-in-memory subarrays,'' \textit{IEEE Trans. Circuits Syst. I, Reg. Papers}, vol. 70, no. 1, pp. 353--363, Jan. 2023.
\bibitem{guo_343_2024} A. Guo, et al., ``34.3 A 22nm 64kb lightning-like hybrid computing-in-memory macro with a compressed adder tree and analog-storage quantizers for transformer and CNNs,'' in \textit{IEEE Int. Solid-State Circuits Conf. (ISSCC)}, Feb. 2024, vol. 67, pp. 570--572.
\bibitem{imagine} A. Kneip et al., "IMAGINE: An 8-to-1b 22nm FD-SOI Compute-In-Memory CfNN Accelerator With an End-to-End Analog Charge-Based 0.15-8POPS/W Macro Featuring Distribution-Aware Data Reshaping," \textit{IEEE Trans. Circuits Syst. for AI}, vol. 2, no. 3, pp. 222-235, Sept. 2025.
\bibitem{aacim} R. Wan et al., "AACIM: A 2785-TOPS/W, 161-TOP/mm2, <1.17\%-RMSE, Analog-In Analog-Out Computing-In-Memory Macro in 28nm," \textit{2024 IEEE European Solid-State Electronics Research Conference (ESSERC)}, Bruges, Belgium, 2024, pp. 581-584.
\bibitem{sehoon_kim_i-bert_2021} S. Kim, et al., ``I-bert: Integer-only bert quantization,'' in \textit{Proc. Int. Conf. Mach. Learn. (ICML)}, Jul. 2021, pp. 5506--5518.
\bibitem{zhikai_li_i-vit_2022} Z. Li, et al., ``I-ViT: Integer-only quantization for efficient vision transformer inference,'' in \textit{Proc. IEEE/CVF Int. Conf. Comput. Vis. (ICCV)}, Oct. 2022, pp. 7698--7708. 
\bibitem{8702715} X. Peng, et al., ``Optimizing weight mapping and data flow for convolutional neural networks on RRAM based processing-in-memory architecture,'' in \textit{Proc. IEEE Int. Symp. Circuits Syst.}, May 2019, pp. 1--5.
\bibitem{xiaochen_peng_heterogeneous_2021} X. Peng, et al., ``Heterogeneous 3-D integration of multitier compute-in-memory accelerators: An electrical-thermal co-design,'' \textit{IEEE Trans. Electron Devices}, vol. 68, no. 9, pp. 4713--4721, Sep. 2021.
\bibitem{lu_runtime_2021} A. Lu et al., ``A runtime reconfigurable design of compute-in-memory-based hardware accelerator for deep learning inference,'' \textit{ACM Trans. Design Autom. Electron. Syst.}, vol. 26, no. 6, Art. no. 45, Jun. 2021.
\bibitem{lee_neurosim_2024} J. Lee, et al., ``NeuroSim V1.4: Extending technology support for digital compute-in-memory toward 1nm node,'' \textit{IEEE Trans. Circuits Syst. I, Reg. Papers}, vol. 71, no. 4, pp. 1733--1744, Apr. 2024.
\bibitem{peng_dnnneurosim_2021} X. Peng, et al., ``DNN+NeuroSim V2.0: An end-to-end benchmarking framework for compute-in-memory accelerators for on-chip training,'' \textit{IEEE Trans. Comput.-Aided Design Integr. Circuits Syst.}, vol. 40, no. 11, pp. 2306--2319, Nov. 2021.
\bibitem{manley_cooptimization_2025} M. Manley, et al., ``Co-optimization of power delivery network design for 3D heterogeneous integration of RRAM-based compute in-memory accelerators,'' \textit{IEEE J. Explor. Solid-State Comput. Devices Circuits}.
\bibitem{li_h3datten_2023} W. Li, et al., ``H3DAtten: Heterogeneous 3-D integrated hybrid analog and digital compute-in-memory accelerator for vision transformer self-attention,'' \textit{IEEE Trans. Very Large Scale Integr. (VLSI) Syst.}, vol. 31, no. 10, pp. 1592--1602, Oct. 2023.
\bibitem{wu_wage_2018} S. Wu, et al., ``Training and inference with integers in deep neural networks,'' arXiv:1802.04680, Feb. 2018.
\bibitem{luo_design_2021} Y.-C. Luo, et al., ``Design of non-volatile capacitive crossbar array for in-memory computing,'' in \textit{IEEE Int. Memory Workshop}, May 2021, pp. 1--4.
\bibitem{10.1145/3649219} Y. Luo, et al., ``H3D-Transformer: A heterogeneous 3D (H3D) computing platform for transformer model acceleration on edge devices,'' \textit{ACM Trans. Design Autom. Electron. Syst.}, vol. 29, no. 3, Apr. 2024, no. 65.
\bibitem{zhou2016dorefa} S. Zhou, et al., ``Dorefa-net: training low bitwidth convolutional neural networks with low bitwidth gradients,'' arXiv:1606.06160, Jun. 2016.
\bibitem{he2019noise} Z. He, et al., ``Noise injection adaption: End-to-end reram crossbar non-ideal effect adaption for neural network mapping,'' in \textit{Proc. Design Autom. Conf. (DAC)}, Jun. 2019, pp. 1--6.
\bibitem{gallo_using_2023} M. L. Gallo, et al., ``Using the IBM analog in-memory hardware acceleration kit for neural network training and inference,'' \textit{APL Mach. Learn.}, vol. 1, no. 4, Dec. 2023, Art. no. 041101.
\bibitem{zhang_asim_2025} J. Zhang, et al., ``ASiM: Modeling and analyzing inference accuracy of SRAM-based analog CiM circuits,'' \textit{IEEE Trans. Circuits Syst. I, Reg. Papers}, 2025.
\bibitem{afm} J. Büchel et al., "Analog foundation models," \textit{arXiv preprint}, 2025.
\bibitem{nora} Y. Hou et al., "NORA: Noise-optimized rescaling of LLMs on analog compute-in-memory accelerators," \textit{IEEE Trans. on Computer-Aided Design of Int. Circuits and Systems}, 2025.
\bibitem{halora} T. Wu et al., "HaLoRA: Hardware-aware Low-Rank Adaptation for Large Language Models Based on Hybrid Compute-in-Memory Architecture," \textit{arXiv preprint}, 2025.
\bibitem{liu2021swin} Z. Liu, et al., "Swin transformer: Hierarchical vision transformer using shifted windows," \textit{Proceedings of the IEEE/CVF international conference on computer vision}, 2021, pp. 10012--10022.
\bibitem{sharda_accelerator_2024} J. Sharda, et al., ``Accelerator design using 3D stacked capacitorless DRAM for large language models,'' in \textit{Proc. IEEE Int. Conf. Artif. Intell. Circuits Syst. (AICAS)}, Apr. 2024, pp. 487--491.
\bibitem{hsu_dram_2025} P.-K. Hsu, et al., ``Monolithic 3D stackable DRAM,'' \textit{IEEE Nanotechnol. Mag.}, vol. 19, no. 2, pp. 7-16, Apr. 2025.

\end{thebibliography}
\end{document}